\documentclass[12pt,onecolumn,twoside]{IEEEtran}

\usepackage{calc}

\usepackage{multirow}
\usepackage{cite}
\usepackage{graphicx,subfigure}
\usepackage{psfrag}
\usepackage{amsmath,amssymb}
\usepackage{color}

\newcommand{\diag}{\mathrm{diag}}

\newcommand{\rank}{\mathrm{rank}}

\DeclareMathOperator*{\dotleq}{\overset{.}{\leq}}
\DeclareMathOperator*{\dotgeq}{\overset{.}{\geq}}
\DeclareMathOperator*{\defeq}{\triangleq}

\newtheorem{theorem}{Theorem}
\newtheorem{corollary}{Corollary}[theorem]

\newtheorem{lemma}{Lemma}

\newtheorem{proposition}{Proposition}
\newtheorem{example}{Example} 

\newcommand{\bit}{\begin{itemize}}
\newcommand{\eit}{\end{itemize}}

\newcommand{\bc}{\begin{center}}
\newcommand{\ec}{\end{center}}

\newcommand{\ba}{\begin{array}}
\newcommand{\ea}{\end{array}}

\newcommand{\beq}{\begin{equation}}
\newcommand{\eeq}{\end{equation}}

\newcommand{\beqn}{\begin{equation*}}
\newcommand{\eeqn}{\end{equation*}}

\newcommand{\bean}{\begin{eqnarray*}}
\newcommand{\eean}{\end{eqnarray*}}
\newcommand{\bea}{\begin{eqnarray}}
\newcommand{\eea}{\end{eqnarray}}

\def\Z{\mathbb{Z}}

\def\R{\mathbb{R}}

\def\E{\mathbb{E}}

\def\av{\boldsymbol{a}}

\def\hv{\boldsymbol{h}}

\def\xv{\boldsymbol{x}}
\def\yv{\boldsymbol{y}}
\def\zv{\boldsymbol{z}}
\def\Am{\boldsymbol{A}}
\def\Bm{\boldsymbol{B}}
\def\Cm{\boldsymbol{C}}
\def\Dm{\boldsymbol{D}}

\def\Gm{\boldsymbol{G}}
\def\Hm{\boldsymbol{H}}

\def\Mm{\boldsymbol{M}}
\def\Nm{\boldsymbol{N}}

\def\Pm{\boldsymbol{P}}
\def\Qm{\boldsymbol{Q}}
\def\Rm{\boldsymbol{R}}
\def\Sm{\boldsymbol{S}}
\def\Tm{\boldsymbol{T}}
\def\Um{\boldsymbol{U}}
\def\Vm{\boldsymbol{V}}

\newcommand{\Dc}{{\mathcal D}}

\newcommand{\Uc}{{\mathcal U}}

\newcommand{\T}{{\scriptscriptstyle\mathsf{T}}}
\renewcommand{\H}{{\scriptscriptstyle\mathsf{H}}}

\newcommand{\CC}{\mathbb{C}}

\newtheorem{remark}{Remark}

\newcommand{\Sigmam}{\pmb{\Sigma}}
\newcommand{\Lambdam}{\pmb{\Lambda}}
\newcommand{\Phim}{\pmb{\Phi}}
\newcommand{\Psim}{\pmb{\Psi}}
\renewcommand{\Bmatrix}[1]{\begin{bmatrix}#1\end{bmatrix}}

\newcommand{\snr}{\mathsf{snr}}
\newcommand{\cond}{\,\vert\,}
\usepackage{mathtools}
\usepackage{xifthen}

\newcommand{\akt}[1][]{\ifthenelse{\isempty{#1}}{\alpha_{k,t}}{\alpha_{#1,t}}}
\newcommand{\aavg}[1][]{\ifthenelse{\isempty{#1}}{\bar{\alpha}}{\bar{\alpha}_{#1}}}
\newcommand{\FBfracP}[1][]{\ifthenelse{\isempty{#1}}{\delta_{\textrm{P}}}{\delta_{\textrm{P},#1}}}
\newcommand{\FBfracPv}[1][]{\ifthenelse{\isempty{#1}}{\boldsymbol{\delta}_{\textrm{P}}}{\boldsymbol{\delta}_{\textrm{P},#1}}}
\newcommand{\FBfracD}[1][]{\ifthenelse{\isempty{#1}}{\delta_{\textrm{D}}}{\delta_{\textrm{D},#1}}}
\newcommand{\CostC}{\mathsf{C}_{\textrm{C}}}
\newcommand{\CostP}{\mathsf{C}_{\textrm{P}}}
\newcommand{\dMAT}{d_{\textrm{MAT}}}
\newcommand{\dsum}{d_{\Sigma}}

\begin{document}
\sloppy
\title{On the Fundamental Feedback-vs-Performance Tradeoff over the MISO-BC with Imperfect and Delayed CSIT}
\author{Jinyuan Chen, Sheng Yang, and Petros Elia
\thanks{An initial version of this paper has been reported as Research Report No. RR-12-275 at EURECOM, December 7, 2012 (see in \cite{CYE:12}). }
\thanks{This paper was submitted in part to the ISIT 2013.}
\thanks{J. Chen and P. Elia are with the Mobile Communications Department, EURECOM, Sophia Antipolis, France (email: \{chenji, elia\}@eurecom.fr). S. Yang is with the Telecommunications department
of SUPELEC, 3 rue Joliot-Curie, 91190 Gif-sur-Yvette, France (e-mail:
sheng.yang@supelec.fr). }
\thanks{The research leading to these results has received funding from the European Research Council under the European Community's Seventh Framework Programme (FP7/2007-2013) / ERC grant agreement no. 257616 (CONECT), from the FP7 CELTIC SPECTRA project, and from Agence Nationale de la Recherche project ANR-IMAGENET.
}
}

\maketitle
\thispagestyle{empty}

\begin{abstract}
This work considers the multiuser multiple-input single-output (MISO) broadcast channel (BC), where a transmitter with $M$ antennas transmits information to $K$ single-antenna users, and where - as expected - the quality and timeliness of channel state information at the transmitter (CSIT) is imperfect.  Motivated by the fundamental question of how much feedback is necessary to achieve a certain performance, this work seeks to establish bounds on the tradeoff between degrees-of-freedom (DoF) performance and CSIT feedback quality.
Specifically, this work provides a novel DoF region outer bound for the general
$K$-user $M\times 1$ MISO BC with partial current CSIT, which naturally bridges the gap between the case
of having no current CSIT (only delayed CSIT, or no CSIT) and the case
with full CSIT.
The work then characterizes the minimum CSIT feedback that is necessary for any point of the sum DoF, which is optimal for the case with $M\geq K$, and the case with $M=2,\ K=3$.  
\end{abstract}

\section{Introduction}

We consider the multiuser multiple-input single-output (MISO) broadcast channel (BC), where a transmitter with $M$ antennas, transmits information to $K$ single-antenna users. In this setting, the received signal at time $t$, is of the form
\begin{align}
y_{k,t} &= \hv_{k,t}^\T \xv_{t} + z_{k,t}, \quad k=1,\cdots,K      \label{eq:chanmodel}
\end{align}
where $\hv_{k,t}$ denotes the $M\times 1$ channel vector for user~$k$, $z_{k,t}$ denotes the unit power AWGN noise, and where $\xv_{t}$ denotes the transmitted signal vector adhering to a power constraint $\E[ ||\xv_{t}||^2 ] \le P$, for $P$ taking the role of the signal-to-noise ratio (snr).
We here consider that the fading coefficients $\hv_{k,t}, \ k=1,\cdots,K$, are independent and identically distributed (i.i.d.) complex Gaussian random variables with zero mean and unit variance, and are i.i.d. over time.

\begin{figure}
	\centering
	\includegraphics[width = 9cm]{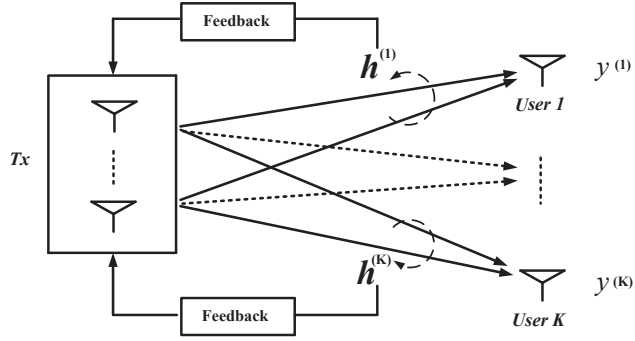}
	\caption{System model of $K$-user MISO BC with CSIT feedback.}
	\label{fig:KuserMISOBCSysModel}
\end{figure}

It is well known that the performance of the BC is greatly affected by the timeliness and quality of feedback; having full CSIT allows for the optimal $\min\{M,K\}$ sum degrees-of-freedom (DoF) (cf.~\cite{CS:03})\footnote{We remind the reader that for an achievable rate tuple $(R_1,R_2,\cdots,R_K)$, where $R_i$ is for user~$i$, the corresponding DoF tuple $(d_1,d_2,\cdots,d_K)$ is given by $d_i = \lim_{P \to \infty} \frac{R_i}{\log P},\ i=1,2,\cdots,K$.  The corresponding DoF region $\Dc$ is then the set of all achievable DoF tuples $(d_1,d_2,\cdots,d_K)$.}, while the absence of any CSIT reduces this to just $1$ sum DoF (cf.~\cite{JG:05,HJSV:12}). This gap has spurred a plethora of works that seek to analyze and optimize BC communications in the presence of delayed and imperfect feedback.  One of the works that stands out is the work by Maddah-Ali and Tse \cite{MAT:11c} which recently revealed the benefits of employing delayed CSIT over the BC, even if this CSIT is completely obsolete. Several interesting generalizations followed, including the work in~\cite{LH:12} which showed that in the BC setting with $K=M+1$, combining delayed CSIT with perfect (current) CSIT (over the last $\frac{K-1}{K}$ fraction of communication period) allows for the optimal sum DoF $M$ corresponding to full CSIT.  
A similar approach was exploited in \cite{TJS:12} which revealed that, to achieve the maximum sum DoF $\min\{M,K\}$, each user has to symmetrically feed back perfect CSIT over a $\frac{\min\{M,K\}}{K}$ fraction of the communication time, and that this fraction is optimal.
Other interesting works in the context of utilizing delayed and current CSIT, can be found in \cite{KYGY:12o,YKGY:12d,GJ:12o,CE:12d} which explored the setting of combining perfect delayed CSIT with immediately available imperfect CSIT, the work in \cite{CE:12c,CE:12it} which additionally considered the effects of the quality of delayed CSIT, the work in \cite{TJSP:12} which considered alternating CSIT feedback, the work in \cite{CE:12m} which considered delayed and progressively evolving (progressively improving) current CSIT, and the works in \cite{VV:11t,GMK:11o,AGK:11o,GMK:11i,XAJ:11b,LSW:12,TMT+:12} and many other publications. 

Our work here generalizes many of the above settings, and seeks to establish fundamental tradeoff between DoF performance and CSIT feedback quality, over the general $K$-user $M\times 1$ MISO BC.

\subsection{CSIT quantification and feedback model}

We proceed to describe the quality and timeliness measure of CSIT feedback, and how this measure relates to existing work.
We here use $\hat{\hv}_{k,t}$ to denote the current channel estimate
(for channel $\hv_{k,t}$) at the transmitter at timeslot $t$, and use
\[\tilde{\hv}_{k,t}= \hv_{k,t} - \hat{\hv}_{k,t}\] to denote the
estimate error assumed to be mutually independent of $\hat{\hv}_{k,t}$
and assumed to have i.i.d. Gaussian entries with power
\[\E\bigl[\|\tilde{\hv}_{k,t}\|^2\bigr]  \doteq P^{-\akt},\]
for some CSI quality exponent $\akt  \in [0,1]$ describing the quality of this estimate. 
We note that  $\akt  = 0$ implies very little current CSIT knowledge,
and that $\akt = 1 $ implies perfect CSIT in terms of the DoF performance\footnote{This can be readily derived, using for example the work in \cite{Caire+:10m}.}.

The approach extends over non-alternating CSIT settings in
\cite{MAT:11c} and \cite{KYGY:12o,YKGY:12d,GJ:12o,CE:12d}, as well as
over an alternating CSIT setting (cf. \cite{TJSP:12,TJS:12}) where CSIT
knowledge alternates between perfect CSIT ($\akt =1$), and delayed or no
CSIT ($\akt =0$).

In a setting where communication takes place over $n$ such coherence periods ($t=1,2,\cdots,n$), this approach offers a natural measure of a per-user average feedback cost, in the form of 
\[\aavg[k] \defeq \frac{1}{n}\sum_{t=1}^n \akt, \quad k=1,2,\cdots,K,\]
as well as a measure of current CSIT feedback cost 
\begin{align}
   \CostC \defeq \sum_{k=1}^{K} \aavg[k],
\end{align}%
accumulated over all users.

\subsubsection{Alternating CSIT setting}
In a setting where delayed CSIT is always available, the above model
captures the alternating CSIT setting where the exponents are binary
($\akt = 0,1$), in which case \[\aavg[k] = \FBfracP[k] \] simply
describes the fraction of time during which user $k$ feeds back perfect
CSIT, with \[\CostC=\CostP \defeq \sum_{k=1}^{K} \FBfracP[k] \] describing the \emph{total perfect CSIT feedback cost}.

\subsubsection{Symmetric and asymmetric CSIT feedback}
Motivated by the fact that different users might have different feedback
capabilities due to the feedback channels with different capacities and
different reliabilities, symmetric CSIT feedback
($\aavg[1]=\cdots=\aavg[K]$) and asymmetric CSIT feedback ($\aavg[k]
\neq \aavg[k'] \ \forall\, k \neq k^{'}$) are considered in this work.

\subsection{Structure of the paper and Summary of Contributions}
Section~\ref{sec:bc-dof} provides the main results of this work:
\bit
\item In Theorem~\ref{thm:con4DoFgeneralMK} we first provide a novel
  outer bound on the DoF region, for the $K$-user $M\times 1$ MISO BC
  with partial current CSIT quantized with $\{\akt\}_{k,t}$, which bridges the case with no current CSIT (only delayed CSIT, or no CSIT) and the case with full CSIT. 
This result	manages to generalize the results by Maddah-Ali and Tse ($\akt = 0, \ \forall t, k$), Yang et al. and Gou and Jafar ($K=2$,
$\akt =\alpha , \ \forall t, k$), Maleki et al. ($K=2$, $\akt[1] = 1,
\akt[2] = 0, \ \forall t$), Chen and Elia ($K=2$, $\akt[1] \neq \akt[2],
\ \forall t $), Lee and Heath ($M=K+1$, $\akt \in\{0,1\},\ \forall t,
k$), and Tandon et al. ($\akt \in\{0,1\},\ \forall t, k$).
\item From Theorem~\ref{thm:con4DoFgeneralMK}, we then provide the upper bound on the sum DoF, which is tight for the case with $M\geq K$ (cf. Theorem~\ref{thm:DoFMgeqK}) and the case with $M=2, K=3$ (cf. Theorem~\ref{thm:FBcostOpM2K3}, Corollary~\ref{cor:FBcostOpM2K3}).
\item Furthermore, Theorem~\ref{thm:FBcostOp} characterizes the minimum
  total current CSIT feedback cost $\CostP^\star$ to achieve the maximum sum DoF, where the total feedback cost $\CostP^\star$ can be distributed among all the users with any (asymmetric and symmetric) combinations $\{
\FBfracP[k]\}_k$.
\item In addition, the work considers some other general settings of BC and provides the DoF inner bound as a function of the CSIT feedback cost.
\eit

The main converse proof, that is for Theorem~\ref{thm:con4DoFgeneralMK}, is shown in the Section~\ref{sec:con4DoFgeneralMK} and appendix. Most of the achievability proofs are shown in the Section~\ref{sec:bc-achiproof}.
Finally Section~\ref{sec:conclu} concludes the paper.

\subsection{Notation and conventions}

Throughout this paper, we will consider communication over $n$ coherence periods where, for clarity of notation, we will focus on the case where we employ a single channel use per such coherence period (unit coherence period). Furthermore, unless stated otherwise, we assume perfect delayed CSIT, as well as adhere to the common convention (see \cite{MAT:11c,MJS:12,GJ:12o,YKGY:12d,TJSP:12,TJS:12}), and assume perfect and global knowledge of channel state information at the receivers.

In terms of notation, $(\bullet)^\T$, $(\bullet)^{\H}$, $\text{tr}(\bullet)$ and $||\bullet||_{F}$ denote the transpose, conjugate transpose, trace and Frobenius norm of a matrix respectively, while $\diag(\bullet)$ denotes a diagonal matrix, $||\bullet||$ denotes the Euclidean norm, and $|\bullet|$ denotes either the magnitude of a scalar or the cardinality of a set.
$o(\bullet)$ and $O(\bullet)$ come from the standard Landau notation, where $f(x) = o(g(x))$ implies $\lim_{x\to \infty} f(x)/g(x)=0$. 
with $f(x) = O(g(x))$ implying that $\limsup_{x\to \infty} |f(x)/g(x)| < \infty$.
We also use $\doteq$ to denote \emph{exponential equality}, i.e., we write $f(P)\doteq P^{B}$ to denote $\displaystyle\lim_{P\to\infty}\frac{\log f(P)}{\log P}=B$. Similarly $\dotgeq$ and $\dotleq$ denote exponential inequalities. 
We use $\Am \succeq \mathbf{0}$ to denote that $\Am $ is positive semidefinite, and use $\Am \preceq \Bm $ to mean that  $\Bm- \Am \succeq \mathbf{0}$.
Logarithms are of base~$2$.

\section{Main results \label{sec:bc-dof}}

\subsection{Outer bounds}

We first present the DoF region outer bound for the general $K$-user $M \times 1$ MISO BC.

\vspace{3pt}
\begin{theorem} [DoF region outer bound]\label{thm:con4DoFgeneralMK}
The DoF region of the $K$-user $M \times 1$ MISO BC, is outer bounded as
\begin{align}
\sum_{k=1}^{K}\! \frac{d_{\pi(k)}}{\min\{k,M\}} &\!\leq\!  1\!+\!
\sum_{k=1}^{K-1}
\!\left(\frac{1}{\min\{k,M\}}\!-\!\frac{1}{\min\{K,M\}}\right)\aavg[\pi(k)] \label{eq:con4DoFgeneralMK}\\
 d_k &\leq  1, \quad k=1,2,\cdots,K
 \end{align}
where $\pi$ denotes a permutation of the ordered set $\{1,2,\cdots,K\}$,
and $\pi(k)$ denotes the $k$~th element of set $\pi$.
\end{theorem}
\vspace{3pt}
\begin{proof}
The proof is shown in Section~\ref{sec:con4DoFgeneralMK}.
\end{proof}

\begin{remark}
It is noted that the bound captures the results in \cite{MAT:11c} ($\akt
= 0, \ \forall t, k$), in \cite{YKGY:12d,GJ:12o} ($K=2$, $\akt =\alpha ,
\ \forall t, k$), in \cite{MJS:12} ($M=K=2$, $\akt[1] = 1,  \akt[2] = 0,
\ \forall\, t$), in \cite{CE:12d} ($K=2$, $\akt[1] \neq \akt[2], \
\forall\,t $), in \cite{TJSP:12,TJS:12} ($\akt \in\{0,1\},\ \forall t, k$), as well as in \cite{KYG:13} ($\alpha^{(k)}_t =\alpha , \ \forall t, k$).
\end{remark}

Summing up the $K$ different bounds from the above, we directly have the
following upper bound on the sum DoF $\dsum \defeq \sum_{k=1}^K d_k$, which is presented using the following notation
\begin{align}
\dMAT &\defeq \frac{K}{1+\frac{1}{\min\{2,M\}}+\frac{1}{\min\{3,M\}}+\cdots+\frac{1}{\min\{K,M\}}} \label{eq:defLambda}\\
\Gamma &\defeq \frac{M}{ \sum_{i=1}^{K-M} \frac{1}{i}(\frac{M-1}{M})^{i-1}+ (\frac{M-1}{M})^{K-M} (\sum_{i=K-M+1}^{K} \frac{1}{i})}. \label{eq:defGamma}
\end{align}

\vspace{3pt}
\begin{corollary} [Sum DoF outer bound] \label{cor:con4DoFgeneralMKsumDoF}
For the $K$-user $M \times 1$ MISO BC, the sum DoF is outer bounded as
\begin{align}
\dsum  \leq \dMAT  + \left(1 - \frac{\dMAT
}{\min\{K,M\}}\right)\sum_{k=1}^{K}\aavg[k].  \label{eq:generalMKsumDoF}
 \end{align}
\end{corollary}
\vspace{3pt}

The above then readily translates onto a lower bound on the minimum
possible total current CSIT feedback cost $\CostC=
\sum_{k=1}^{K}\aavg[k]$ needed to achieve the maximum sum
DoF\footnote{Naturally the result is limited to the case where
$\min\{K,M\}>1$.} $\dsum =\min\{K,M\}$.

\vspace{3pt}
\begin{corollary} [Bound on CSIT cost for maximum DoF] \label{cor:con4minsumcost}
The minimum $\CostC$ required to achieve the maximum sum DoF $\min\{K,M\}$ of the $K$-user $M \times 1$ MISO BC, is lower bounded as 
\begin{align}
\CostC^{\star} \geq  \min\{K,M\}.
 \end{align}
\end{corollary}
\vspace{3pt}

Transitioning to the alternating CSIT setting where $\akt \in \{0, 1\}$,
we have the following sum-DoF outer bound as a function of the
perfect-CSIT duration $\aavg[k] = \FBfracP[k] = \FBfracP, \ \forall \,
k$.  We note that the bound holds irrespective of whether, in the
remaining fraction of the time $1-\FBfracP$, the CSIT is delayed or non existent.

\vspace{3pt}
\begin{corollary} [Outer bound, alternating CSIT]\label{cor:con4DoFgeneralMK}
For the $K$-user $M \times 1$ MISO BC, the sum DoF is outer bounded as
\begin{align}
\dsum  \leq \dMAT  + \left(K - \frac{K\dMAT
}{\min\{K,M\}}\right)\min\left\{\FBfracP, \frac{\min\{K,M\}}{K}\right\}.
 \end{align}
\end{corollary}

\subsection{Optimal cases of DoF characterizations}

We now provide the optimal cases of DoF characterizations. The case with $M\geq K$ is first considered in the following.
\vspace{2pt}
\begin{theorem} [Optimal case, $M\geq K$] \label{thm:DoFMgeqK}
For the $K$-user $M\times 1$ MISO BC with $M\geq K$, the optimal sum DoF is characterized as
  \begin{align}
	   \dsum  = (K - \dMAT)\min\{\FBfracP, 1\}  +\dMAT.
  \end{align}
\end{theorem}
\vspace{2pt}
\begin{proof}
The  converse and achievability proofs are derived from Corollary~\ref{cor:con4DoFgeneralMK} and Proposition~\ref{pr:DoFMgeq2P} (shown in the next subsection), respectively.
\end{proof}

\begin{remark}
It is noted that, for the special case with $M=K=2$, the above characterization captures the result in \cite{TJSP:12}.
\end{remark}

\begin{figure}
	\centering
	\includegraphics[width = 8cm]{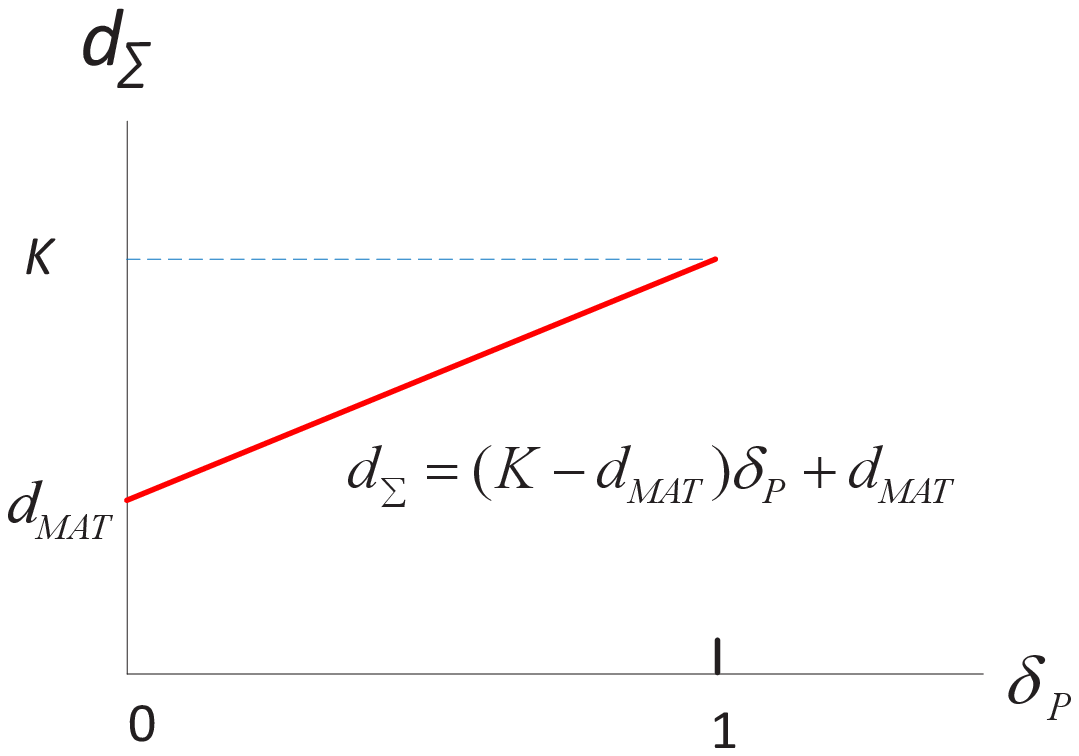}
	\caption{Optimal sum DoF $\dsum $ vs. $\FBfracP$ for the MISO BC with $M\geq K$ .}
	\label{fig:SumDoF_P_Mgeq2KleqM}
\end{figure}

Moving to the case where $M < K$, we have the following optimal sum DoF
characterizations for the case with $M=2,\ K=3$. The first interest is
placed on the minimum $\CostP^{\star}(\dsum )$ to achieve a sum DoF
$\dsum$, recalling that $\CostP^{\star} = \sum_{k=1}^{K} \FBfracP[k]$ describes the total perfect CSIT feedback cost.
\vspace{3pt}
\begin{theorem} [Optimal case, $M=2, K=3$] \label{thm:FBcostOpM2K3}
For the three-user $2\times 1$ MISO BC, the minimum  total perfect CSIT feedback cost is characterized as
\begin{align}
  \CostP^{\star}(\dsum) = (4\dsum-6 )^{+}, \quad  \forall \  \dsum \in [0,2]
\end{align}
where the total feedback cost $\CostP^{\star}(\dsum)$ can be
distributed among all the users with some combinations $\{
\FBfracP[k]\}_k$ such that $\FBfracP[k] \leq
\CostP^{\star}(\dsum)/2$ for any $k$.
\end{theorem}
\vspace{3pt}
\begin{proof}
The converse proof is directly from Corollary~\ref{cor:con4DoFgeneralMKsumDoF}, while the achievability proof is shown in Section~\ref{sec:poofFBM2K3Ach}.
\end{proof}

\begin{figure}
	\centering
	\includegraphics[width = 8cm]{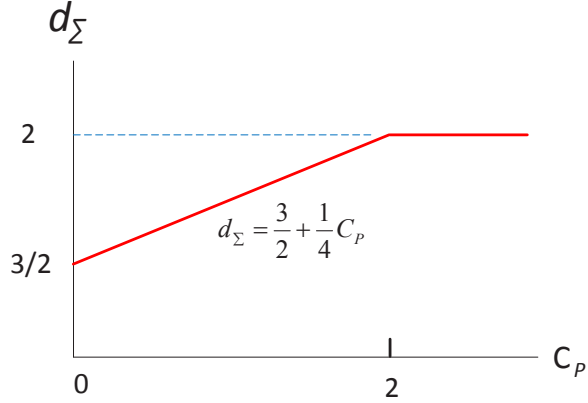}
	\caption{Optimal sum DoF ($\dsum$) vs. total perfect CSIT
        feedback cost ($\CostP$) for three-user $2\times 1$ MISO BC.}
	\label{fig:FBcostOpM2K3}
\end{figure}

Theorem~\ref{thm:FBcostOpM2K3} reveals the fundamental tradeoff between sum DoF and total perfect CSIT feedback cost (see Fig~\ref{fig:FBcostOpM2K3}). The following examples are provided to offer some insights corresponding to Theorem~\ref{thm:FBcostOpM2K3}.

\begin{example}
For the target sum DoF $\dsum= 3/2, \ 7/4, \  2$, the minimum total
perfect CSIT feedback cost is $\CostP^\star = 0, \ 1,\ 2$, respectively.
\end{example}

\begin{example}
The target $\dsum=7/4$ is achievable with asymmetric feedback
$\FBfracPv=[1/6\quad 1/3\quad 1/2]$, and
symmetric feedback $\FBfracPv=[1/3\quad 1/3\quad 1/3]$, and some other feedback
such that $\CostP^\star(7/4)=1$.
\end{example}

\begin{example}
The target $\dsum=2$ is achievable with asymmetric feedback
$\FBfracPv=[1/3\quad 2/3\quad 1]$,
and symmetric feedback 
$\FBfracPv=[2/3\quad 2/3\quad 2/3]$, and some other feedback
such that $\CostP^\star(2)=2$.
\end{example}

Transitioning to the symmetric setting where $\FBfracP[k] = \FBfracP \
\forall\, k$, from Theorem~\ref{thm:FBcostOpM2K3} we have the
fundamental tradeoff between optimal sum DoF and CSIT feedback cost
$\FBfracP$.
\vspace{3pt}
\begin{corollary} [Optimal case, $M=2, K=3$, $\FBfracP$] \label{cor:FBcostOpM2K3}
For the three-user $2\times 1$ MISO BC with symmetrically alternating CSIT feedback, the optimal sum DoF is characterized as
\begin{align}
\dsum =   \min \left\{\frac{3(2+\FBfracP)}{4}, 2 \right\}.
\end{align}
\end{corollary}
\vspace{3pt}

Now we address the questions of what is the minimum $\CostP^{\star}$ to
achieve the maximum sum DoF $\min\{M,K\}$ for the general BC, and how to
distribute $\CostP^\star$ among all the users, recalling again that
$\CostP^\star$ is the total perfect CSIT feedback cost.
\vspace{3pt}
\begin{theorem} [Minimum cost for maximum DoF] \label{thm:FBcostOp}
For the $K$-user $M\times 1$ MISO BC, the minimum  total perfect CSIT feedback cost to achieve the maximum DoF is characterized as
\begin{eqnarray}
\CostP^\star(\min\{M,K\})  =  \left\{
\begin{array}{l l}
   0  ,        &  \text{if}  \quad \min\{M,K\}=1  \\
  \min\{M,K\} , & \text{if} \quad  \min\{M,K\}>1
\end{array} \right. \nonumber
\end{eqnarray}
where the total feedback cost $\CostP^\star$ can be distributed among
all the users with any combinations $\{ \FBfracP[k] \}_k$.
\end{theorem}
\vspace{3pt}
\begin{proof}
For the case with $\min\{M,K\}=1$, simple TDMA is optimal in terms of the DoF performance. For the case with $\min\{M,K\}>1$, the converse proof is directly derived from Corollary~\ref{cor:con4minsumcost}, while the achievability proof is shown in Section~\ref{sec:poof01}.
\end{proof}

It is noted that Theorem~\ref{thm:FBcostOp} is a generalization of the result in \cite{TJS:12} where only symmetric feedback was considered. 
The following examples are provided to offer some insights corresponding to Theorem~\ref{thm:FBcostOp}.

\begin{example}
For the case where $M=2, \ K=4$,  the optimal 2 sum DoF performance is
achievable, with asymmetric feedback $\FBfracPv = [1/5\quad
2/5\quad3/5\quad4/5]$, and symmetric feedback $\FBfracPv = [1/2
\quad 1/2 \quad 1/2 \quad 1/2]$, and any other feedback such that $\CostP^\star = 2$.
\end{example}
\begin{example}
For the case where $M=3, \ K=5$,  the optimal 3 sum DoF performance is
achievable, with asymmetric feedback $\FBfracPv = [1/5 \quad 2/5 \quad
3/5 \quad 4/5 \quad 1 ]$, and symmetric feedback $\FBfracPv =
[3/5\quad3/5\quad3/5\quad3/5\quad3/5]$, and any other feedback such that $\CostP^\star=3$.
\end{example}

The following corollary is derived from Theorem~\ref{thm:FBcostOp}, where the case with $\min\{M,K\}>1$ is considered.
\begin{corollary} [Minimum cost for maximum DoF] \label{pro:asymoptJ}
For the $K$-user $M\times 1$ MISO BC, where $J$ users instantaneously feed back perfect (current) CSIT, with the other users feeding back delayed CSIT, then the minimum number $J$ is $\min\{M,K\}$, in order to achieve the maximum sum DoF $\min\{M,K\}$.
\end{corollary}
\vspace{3pt}

\subsection{Inner bounds}

In this subsection, we provide the following inner bounds on the sum DoF as a function of the CSIT cost, which are tight for many cases as stated.
\vspace{3pt}
\begin{proposition} [Inner bound, $M=2, K\geq 3$] \label{lm:DoFM2KPD}
For the $K(\geq 3)$-user $2\times 1$ MISO BC, the sum DoF is bounded as 
  \begin{align} \label{eq:DoFM2KPD1}
	   \dsum \geq \frac{3}{2} + \frac{K}{4} \min\{\FBfracP, \frac{2}{K}\}.
  \end{align}
\end{proposition}
\vspace{3pt}
\begin{proof}
The proof is shown in Section~\ref{sec:poof4}.
\end{proof}

\begin{figure}
	\centering
	\includegraphics[width = 8cm]{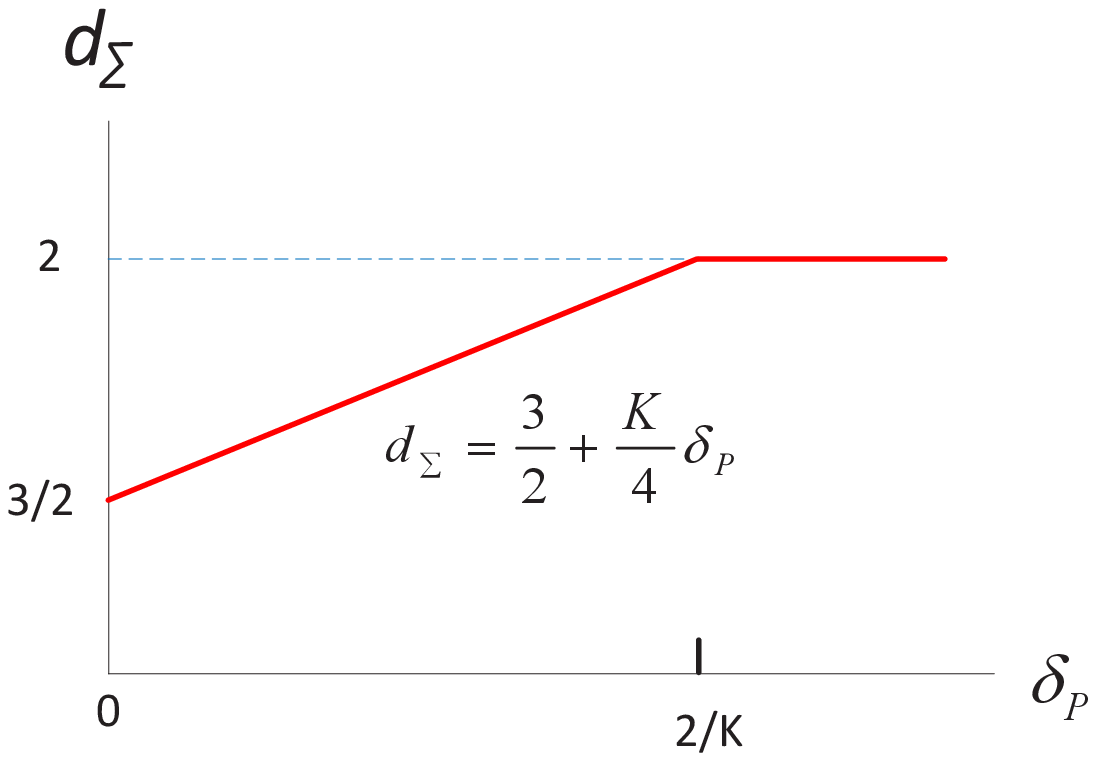}
	\caption{Achievable sum DoF $\dsum$ vs. $\FBfracP$ for the $K (\geq 3)$-user $2\times 1$ MISO BC.}
	\label{fig:SumDoF_P_M2K}
\end{figure}

\vspace{3pt}
\begin{proposition} [Inner bound, $M\geq K$ and $M< K$] \label{pr:DoFMgeq2P}
For the $K$-user $M\times 1$ MISO BC,  the sum DoF for the case with $M\geq K$ is bounded as 
  \begin{align} 
	   \dsum \geq (K - \dMAT)\min\{\FBfracP, 1\}  +\dMAT,
  \end{align}
while for the case with $M<K$, the sum DoF is bounded as 
  \begin{align} 
	   \dsum \geq (K - \frac{K \Gamma }{M}) \min \{\FBfracP, \frac{M}{K} \} +\Gamma.
  \end{align}
\end{proposition}
\vspace{3pt}

\begin{proof}
The proof is shown in Section~\ref{sec:poof5}.
\end{proof}

\begin{figure}
	\centering
	\includegraphics[width = 8cm]{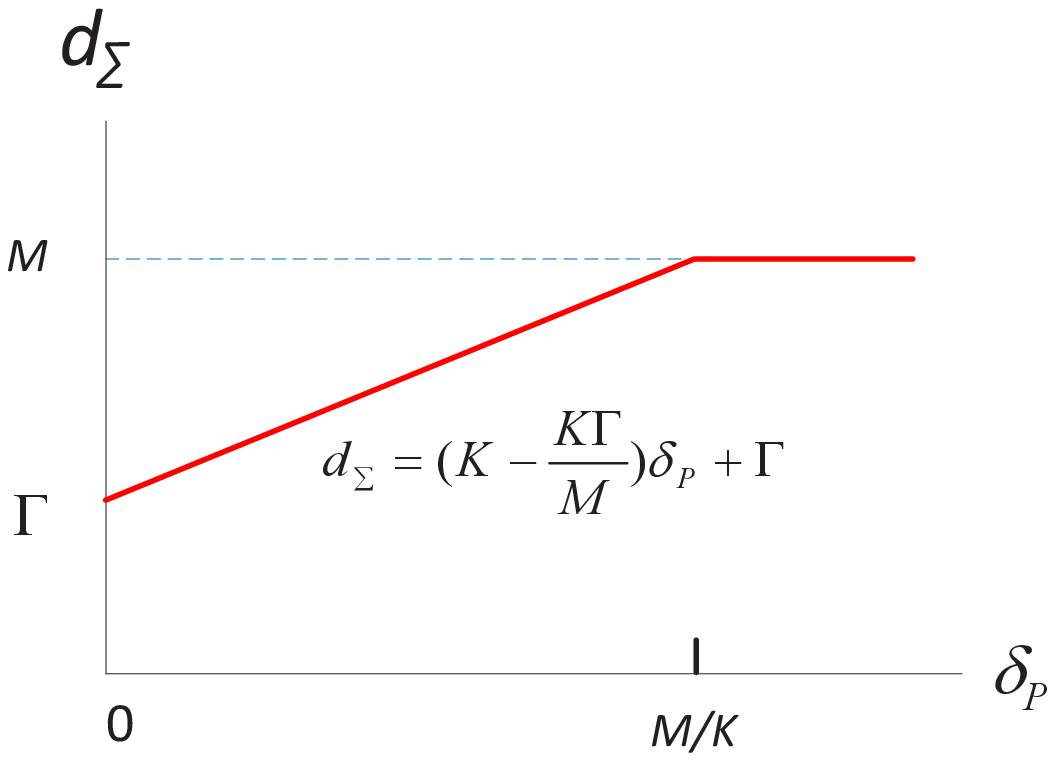}
	\caption{Achievable sum DoF $\dsum$ vs. $\FBfracP$ for the MISO BC with $M<K$.}
	\label{fig:SumDoF_P_Mgeq2KgeqM}
\end{figure}

Finally, we consider a case of BC with delayed CSIT feedback only, where
$\FBfracP=0$. In this case, we use $\FBfracD[k]$
to denote the fraction of time during which CSIT fed back from user $k$
is delayed, and focus on the case with $\FBfracD[k]=\FBfracD, \ \forall k$.

\vspace{3pt}
\begin{proposition} [Inner bound on DoF with delayed CSIT] \label{lm:DoFM2KD}
For the $K (\geq 3)$-user $2\times 1$ MISO BC, and for the case of
$\FBfracP=0$, the sum DoF is bounded as 
  \begin{align} \label{eq:DoFM2KD}
	   \dsum \geq \min \left\{1+\frac{K}{2}\FBfracD, \
           \frac{12}{11}+\frac{4K}{11} \FBfracD, \  \frac{3}{2} \right\}.
  \end{align}
\end{proposition}
\vspace{3pt}
\begin{proof}
The proof is shown in Section~\ref{sec:poof3}.
\end{proof}

\begin{figure}
	\centering
	\includegraphics[width = 8cm]{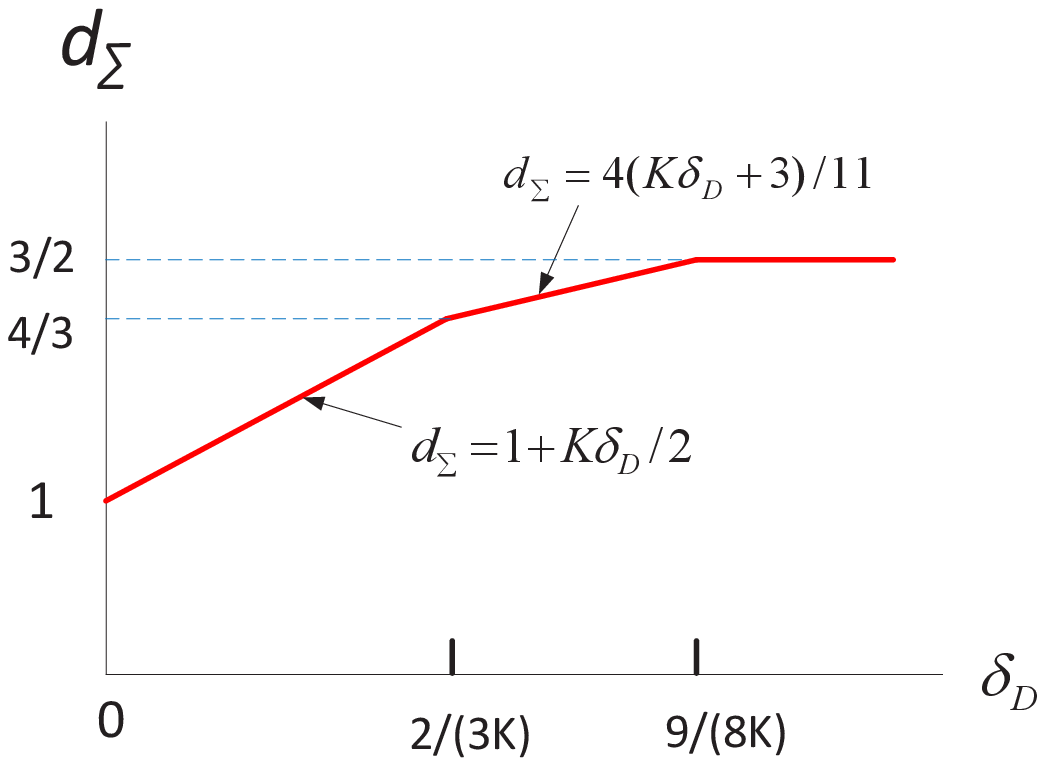}
	\caption{Achievable sum DoF $\dsum$ vs. $\FBfracD$ for the
        MISO BC with $K\ge3, M=2$, where $\FBfracP=0$.}
	\label{fig:SumDoF_D_M2K}
\end{figure}

\begin{remark}
For the $K$-user MISO BC with current and delayed CSIT feedback, by
increasing the number of users, the same DoF performance can be
achievable with decreasing feedback cost per user. For example, for the
$K$-user MISO BC with $M=2$, by increasing $K$ we can achieve any fixed
DoF within the range of $(1, 2]$, with decreasing $\FBfracP \leq
\frac{2}{K}$, and $\FBfracD \leq \frac{9}{8K}$, both of which approach to $0$ as $K$ is large.
\end{remark}

\section{Converse proof of Theorem~\ref{thm:con4DoFgeneralMK} \label{sec:con4DoFgeneralMK}}

We first provide the Proposition~\ref{pro:entropydiff} to be used, where we drop the time index for simplicity. 
\vspace{3pt}
\begin{proposition} \label{pro:entropydiff}
  Let  
  \begin{align*}
	    y_k &= \hv^\T_k \xv + z_k, \\ 
    \yv_k &\triangleq [y_1 \ y_2 \ \cdots \ y_k]^\T        \\
    \zv_k &\triangleq [z_1 \ z_2 \ \cdots \ z_k]^\T        \\
		\Hm_k &\triangleq [\hv_1 \ \hv_2 \ \cdots \ \hv_k]^\T  \\
    \Hm &\triangleq [\hv_1 \ \hv_2 \ \cdots \ \hv_K]^\T  \\
    \Hm &= \hat{\Hm} + \tilde{\Hm}
  \end{align*}%
  where $\tilde{\hv}_i \in \CC^{M\times 1}$ has i.i.d.~$\mathcal{N}_{\mathbb{C}} (0,\sigma^2_i)$ entries.
  Then, for any $U$ such that $p_{X|U\hat{H}\tilde{H}} =p_{X|U\hat{H}}$
and $K\ge m \ge l$, we have 
    \begin{equation}
      l\,'\, h(\yv_{m}|U,\hat{H},\tilde{H}) - m'\,
      h(\yv_l|U,\hat{H},\tilde{H}) 
      \le - (m'-l\,') \sum_{i=1}^l \log \sigma^2_i + o(\log \snr)
      \label{eq:tmp2}
    \end{equation}%
    where we define $l\,' \triangleq \min\left\{ l, M \right\}$ and $m' \triangleq
    \min\left\{ m, M \right\}$.
\end{proposition}
\vspace{3pt}
\begin{proof}
The proof is shown in the Section~\ref{sec:entropydiff}.
\end{proof}

Now giving the observations and messages of users $1,\ldots,k-1$ to user
$k$, we establish the following genie-aided upper bounds on the achievable
rates 
\begin{align}
  n R_1 &\le I(W_1; y_1^n \cond \Omega^n) + n \epsilon \\
  n R_2 &\le I(W_2; y_1^n,y_2^n \cond W_1, \Omega^n) + n \epsilon \\
  &\ \vdots \nonumber \\
  n R_K &\le I(W_K; y_1^n,y_2^n,\ldots,y_K^n \cond
  W_1,\ldots,W_{K-1},\Omega^n) + n \epsilon 
\end{align}%
where we apply Fano's inequality and some basic chain rules of mutual
information using the fact that messages from different users are independent,
where we define
\begin{align*}
\Sm_{t} \defeq & \ \Bmatrix{\hv_{1,t} \ \cdots \ \hv_{K,t}}^\T \\
\hat{\Sm}_{t} \defeq& \ \Bmatrix{\hat{\hv}_{1,t} \ \cdots \ \hat{\hv}_{K,t}}^\T \\
\Omega^{n} \defeq &  \ \{ \Sm_{t}, \hat{\Sm}_{t}\}_{t=1}^{n} \\
y_k^n \defeq &  \  \{y_{k,t}\}_{t=1}^{n}.
\end{align*} 
Alternatively, we have
\begin{align}
  n R_1 &\le h(y_1^n \cond \Omega^n) - h(y_1^n \cond W_1, \Omega^n) + n \epsilon \\
  n R_2 &\le h(y_1^n,y_2^n \cond W_1, \Omega^n) - h(y_1^n,y_2^n \cond
  W_1, W_2, \Omega^n) + n \epsilon \\
  &\ \vdots \nonumber \\
  n R_K &\le h(y_1^n,\ldots,y_K^n \cond W_1,\ldots,W_{K-1}, \Omega^n) -
  h(y_1^n,\ldots,y_K^n \cond W_1,\ldots,W_{K},\Omega^n) + n \epsilon. 
\end{align}
Therefore, it follows that
\begin{align}
  &\sum_{k=1}^K \frac{n}{k'} (R_k-\epsilon) \nonumber \\
  &\le \sum_{k=1}^{K-1} \left( \frac{1}{(k+1)'} h(y_1^n,\ldots,y_{k+1}^n
  \cond W_1,\ldots,W_{k}, \Omega^n) - \frac{1}{k'}
  h(y_1^n,\ldots,y_k^n \cond W_1,\ldots,W_{k}, \Omega^n) \right)
  \nonumber \\
  &\qquad + h(y_1^n \cond \Omega^n) -
  \frac{1}{K'} h(y_1^n,\ldots,y_K^n \cond W_1,\ldots,W_{K},\Omega^n) \\
  &\le \sum_{k=1}^{K-1} \sum_{t=1}^n \biggl(
  \frac{1}{(k+1)'} h(y_{1,t},\ldots,y_{k+1,t} \cond y_1^{t-1},\ldots,y_{k}^{t-1}, 
  W_1,\ldots,W_{k}, \Omega^n) \nonumber \\
  &\qquad - \frac{1}{k'}
  h(y_{1,t},\ldots,y_{k,t} \cond y_1^{t-1},\ldots,y_{k}^{t-1}, 
  W_1,\ldots,W_{k}, \Omega^n) \biggr) + n\log P + n\,o(\log P)
  \label{eq:tmp61} \\
  &\le  \log P \sum_{k=1}^{K-1} \sum_{t=1}^n \frac{(k+1)'-k'}{k'
  (k+1)'}\sum_{i=1}^k \akt[i] + n\log P + n\,o(\log P)
  \label{eq:tmp62} \\
  &=  n \log P \sum_{k=1}^{K-1} \frac{(k+1)'-k'}{k' (k+1)'}\sum_{i=1}^k
  \aavg[i] + n\log P + n\,o(\log P) \\
  &=  n \log P \sum_{k=1}^{K-1} \Bigl(\frac{1}{k'} -  \frac{1}{K'}
  \Bigr) \aavg[k] + n\log P + n\,o(\log P) 
\end{align}%
where we define 
\begin{align}
  k' &\triangleq \min\left\{ k, M \right\};
\end{align}%
the inequality \eqref{eq:tmp61} is due to 1)~the chain rule of differential
entropy, 2)~the fact that removing condition does not decrease differential
entropy, 3)~$h(y_{1,t} \cond \Omega^n)\le \log P + o(\log P)$, i.e.,
Gaussian distribution maximizes differential entropy under covariance
constraint, and 4)~$h(y_1^n,\ldots,y_K^n \cond
W_1,\ldots,W_{K},\Omega^n) = h(z_{1,1},z_{1,2},\ldots,z_{K,n}) > 0$; 
\eqref{eq:tmp62} is from Proposition~\ref{pro:entropydiff} by setting
$U = \{y_1^{t-1},\ldots,y_{k}^{t-1}, W_1,\ldots,W_{k},\Omega^n\}
\setminus \{\Sm_t, \hat{\Sm}_t \}$, ${H} = {\Sm_t}$, and $\hat{H} = \hat{\Sm_t}$; 
the last
equality is obtained after putting the summation over $k$ inside the summation
over $i$ and some basic manipulations. 
Similarly, we can interchange the roles of the users and obtain the same
genie-aided bounds. 
Finally, the single antenna constraint gives that $d_i \leq 1, \ i=1,\cdots,K$.
With this, we complete the proof.

\section{Details of achievability proofs \label{sec:bc-achiproof} }

In this section, we provide the details of the achievability proofs.
Specifically, the achievability proof of Theorem~\ref{thm:FBcostOp} is first described in Section~\ref{sec:poof01}, which can be applied in parts for the achievability proof of Theorem~\ref{sec:poofFBM2K3Ach} shown in Section~\ref{sec:poofFBM2K3Ach}, with the proposition proofs shown in the rest of this section. 

\subsection{Achievability proof of Theorem~\ref{thm:FBcostOp}  \label{sec:poof01}}

We will prove that, the optimal sum DoF $ \dsum =  \min\{M,K\}$ is
achievable with any CSIT feedback cost $\FBfracPv \defeq
[ \FBfracP[1]\quad \FBfracP[2]\quad \cdots\quad \FBfracP[K]]  \in
\R^{K} $ such that $\CostP = \sum_{k=1}^{K} \FBfracP[k] = \min\{M,K\}$.
First of all, we note that there exists a minimum number $n$ such that
\[  \FBfracPv' \defeq   [ \FBfracP[1]'\quad \FBfracP[2]'\ \cdots\ 
\FBfracP[K]'] \defeq n\FBfracPv  =   [n\FBfracP[1]\quad n\FBfracP[2]
\ \cdots\  n\FBfracP[K]  ]  \in \Z^{K}\] is an integer vector.
The explicit communication with $n$ channel uses is given as follows:
\bit
\item Step~1: Initially set time index $t=1$.
\item Step~2: Permute user indices orderly into a set $\Uc$ such that
  $\FBfracP[\Uc(1)]' \leq \FBfracP[\Uc(2)]' \leq \cdots \leq
  \FBfracP[\Uc(K)]'$, where $\Uc(k)$ denotes the $k$~th element of set $\Uc$, and where $\Uc(k) \in \{ 1,2,\cdots,K\}$.
\item Step~3: Select $\min\{M,K\}$ users to communicate: users~$\Uc(K-\min\{M,K\}+1),\cdots, \Uc(K-1), \Uc(K)$.
\item Step~4: Let selected users feed back perfect CSIT at time~$t$, keeping the rest users silent.
\item Step~5: The transmitter sends $\min\{M,K\}$ independent symbols to those selected users respectively, which can be done with simple zero-forcing.
\item Step~6: Set  $\FBfracP[\Uc(k)]' =\FBfracP[\Uc(k)]' -1, \ k=K-\min\{M,K\}+1, \cdots, K-1, K$.
\item Step~7: Set $t=t+1$. If renewed $t > n$ then terminate, else go back to step~2.
\eit
In the above communication with $n$ channel uses, the algorithm
guarantees that user~$i$ is selected by
$\FBfracP[k]'=n\FBfracP[k]$ times totally, and that
$\min\{M,K\}$ different users are selected in each channel use. As a
result, the optimal sum DoF $ \dsum =\min\{M,K\}$ is achievable. 

Now we consider an example with  $M=2, \ K=3$, and
$\FBfracPv = [1/3 \quad 2/3 \quad 1]$,
and show that the optimal sum DoF $ \dsum =2$ is achievable with the following communication:
\bit
\item Let $n=3$. Initially $\FBfracP[1]'=n\FBfracP[1]= 1$,
  $\FBfracP[2]'=n\FBfracP[2]= 2$, $\FBfracP[3]'=n\FBfracP[3]=3$.
\item For $t=1$, we have $\Uc = \{1,2,3\}$, and  $\FBfracP[\Uc(1)]' =1$,
  $\FBfracP[\Uc(2)]' =2$, $\FBfracP[\Uc(3)]' =3$.  Users~3 and~2 are selected to communicate.    
\item For $t=2$, we update the parameters as $\Uc = \{1,2,3\}$, and
  $\FBfracP[\Uc(1)]' =1$, $\FBfracP[\Uc(2)]' =1$, $\FBfracP[\Uc(3)]' =2$.  At this time, again user~3 and user~2 are selected to communicate.
\item For $t=3$, we update the parameters as  $\Uc = \{2,1,3\}$, and
  $\FBfracP[\Uc(1)]' =0$, $\FBfracP[\Uc(2)]' =1$, $\FBfracP[\Uc(3)]' =1$.  At this time, user~3 and user~1 are selected to communicate. After that the communication terminates.
\eit
In the above communication with three channel uses, the transmitter
sends two symbols in each channel use, which allows for the optimal sum
DoF $ \dsum =2$  (see Table~\ref{tab:proof0t1}).

\begin{table}
\caption{Summary of the scheme for achieving $d^{*}_{\sum}=2$ with
$\CostP^\star =2$, where $M=2,\ K=3$,  $\FBfracP[1]=1/3, \
\FBfracP[2]=2/3, \ \FBfracP[3]=1$.}
\begin{center}
\begin{tabular}{|c|c|c|c|c|}
  \hline
time  $t$                &  1              & 2              & 3      \\
   \hline
 $\Uc$                &  $ \{1,2,3\}$  &  $ \{1,2,3\}$               & $\{2,1,3\} $       \\
   \hline
$\{ \FBfracP[\Uc(1)]',\  \FBfracP[\Uc(2)]', \ \FBfracP[\Uc(3)]'\}$  &  $ \{1,2,3\}$  &  $ \{1,1,2\}$               & $\{0,1,1\} $       \\
   \hline
 Active users         & user~2, 3     &  user~2, 3         & user~1, 3  \\
   \hline
	Perfect CSIT feedback   &  user~3: yes  & user~3: yes  & user~3: yes   \\
	                        &  user~2: yes  & user~2: yes  & user~2: no   \\
	                        &  user~1: no   & user~1: no   & user~1: yes \\
   \hline
	No. of transmitted symbols    &  $2$          & $2$          & $2$         \\
	\hline
\end{tabular}
\end{center}
\label{tab:proof0t1}
\end{table}

\subsection{Achievability proof of Theorem~\ref{thm:FBcostOpM2K3}  \label{sec:poofFBM2K3Ach}}

We proceed to show that, any sum DoF $ \dsum \in [3/2, 2]$ is
achievable with the feedback \[\FBfracP[k] \leq  \frac{\CostP}{2}, \
k=1,2,3 ,  \quad \text{such that} \quad \CostP= \sum_{k=1}^{3}
\FBfracP[k] =  4 \dsum -6.\]

First of all, we note that there exists a minimum number $n$ such that
\[[ 2n\FBfracP[1]/\CostP\quad 2n\FBfracP[2]/\CostP\quad
n2\FBfracP[3]/\CostP  ] \in \Z^{3}, \quad \text{and} \quad
2n/\CostP \in \Z. \]
The scheme has two blocks, with the first block consisting of $n$
channel uses, and the second block consisting of \[n^{'}=2n/\CostP -n\] channel uses.
In the first block, we use the algorithm shown in the
Section~\ref{sec:poof01} to achieve the full sum DoF in those $n$
channel uses, during which user~$k$ feeds back perfect CSIT in
$2n\FBfracP[3]/\CostP$ channel uses, for $k=1,2,3$. In the second block, we use the Maddah-Ali and Tse scheme in \cite{MAT:11c} to achieve 3/2 sum DoF in those $n^{'}$ channel uses, during which each user feeds back delayed CSIT only.

The communication with $n$ channel uses for the first block is given as follows:
\bit
\item Step~1: Let $\FBfracP[k]' = 2n\FBfracP[k]/\CostP$ for all $k$. Initially, set $t=1$.
\item The steps~2, 3, 4, 5, 6 are the same as those in the algorithm shown in Section~\ref{sec:poof01}, for $M=2, K=3$.
\item Step~7: Set $t=t+1$. If renewed $t > n$ then terminate, else go back to step~2.
\eit
In the above communication with $n$ channel uses, the algorithm
guarantees that user~$k$, $k=1,2,3,$ is selected by
$\FBfracP[k]'=2n\FBfracP[k]/\CostP$ times. We note that
$\FBfracP[k]' \leq n$ under the constraint $\FBfracP[k] \leq
\CostP /2 $ for any $k$, and that $\sum^{K}_{k=1}\FBfracP[k]' = 2n$, to
suggest that in each timeslot two different users are selected, which
allows for the optimal $2$ sum DoF in this block. 

As stated, in the second block, we use the MAT scheme to achieve the 3/2 sum DoF in those $n^{'}$ channel uses, during which each user feeds back delayed CSIT only. 
As a result, in the total $n+n^{'}$ channel uses communication,
user~$k=1,2,3$ feeds back perfect CSIT in
$2n\FBfracP[k]/(\CostP(n+n^{'})) = \FBfracP[k] $ fraction of communication period, with achievable sum DoF given as 
\[ \dsum = \frac{2n}{(n+n^{'})} + \frac{3n^{'}}{2(n+n^{'})} =
\frac{3}{2}+\frac{1}{4} \CostP .\]

We note that the achievability scheme applies to the case of having some
$\FBfracP[1], \FBfracP[2],\FBfracP[3] \leq  \CostP/2 $ such
that $\CostP =  4 \dsum -6$, and allows to achieve any sum DoF $ \dsum \in [3/2, 2]$. 
Apparently, $\CostP = 0$ allows for any sum DoF $ \dsum \in [0, 3/2]$, which completes the proof. 

\subsection{Proof of Proposition~\ref{lm:DoFM2KPD} \label{sec:poof4}}

The achievability scheme is based on time sharing between two strategies
of CSIT feedback, i.e., delayed CSIT feedback with $\FBfracP'=0$ and
alternating CSIT feedback with $\FBfracP''= \frac{2}{K}$, where the
first strategy achieves  $\dsum'=3/2$ by applying Maddah-Ali and Tse
(MAT) scheme (see in \cite{MAT:11c}), with the second strategy achieving
$\dsum''=2$ by using alternating CSIT feedback manner (see in \cite{TJS:12}).

Let $\Delta \in [0,1]$ (res. $1-\Delta$) be the fraction of time during which the first (res. second) CSIT feedback strategy is used in the communication. As a result, the final feedback cost (per user) is given as 
\begin{align}
  \FBfracP = \FBfracP' \Delta + \FBfracP'' (1-\Delta),  \label{eq:achproof21}
\end{align}%
implying that 
\begin{align}
  \Delta  = \frac{\FBfracP''-\FBfracP}{\FBfracP''-\FBfracP'}, \label{eq:achproof22}
\end{align}%
with final sum DoF given as 
\begin{align}
	\dsum &= \dsum' \Delta + \dsum'' (1-\Delta) \nonumber\\
	&= \dsum'' +   \Delta (\dsum' - \dsum'')   \nonumber\\
  &= \dsum'' +  (\dsum' - \dsum'')
  \frac{\FBfracP''-\FBfracP}{\FBfracP''-\FBfracP'}  \nonumber\\	
  &= \frac{3}{2} + \frac{K}{4} \FBfracP    \label{eq:achproof23}
\end{align}%
which completes the proof.

\subsection{Proof of Proposition~\ref{pr:DoFMgeq2P} \label{sec:poof5}}

For the case with $M\geq K$, the proposed scheme is based on time
sharing between delayed CSIT feedback with $\FBfracP'=0$  and full
CSIT feedback with $\FBfracP''= 1$, where the first feedback
strategy achieves $d^{\ '}_{\sum}=\dMAT$ by applying MAT scheme, with the second one achieving $d^{\ ''}_{\sum}=K$.
As a result, following the steps in \eqref{eq:achproof21}, \eqref{eq:achproof22}, \eqref{eq:achproof23}, the final sum DoF is calculated as 
\begin{align*}
	\dsum &= \dsum'' +  (\dsum' - \dsum'')
        \frac{\FBfracP''-\FBfracP}{\FBfracP''-\FBfracP'}  \\	
  &= (K - \dMAT)\FBfracP  +\dMAT  
\end{align*}%
where  $\FBfracP\in [0, 1]$ is the final feedback cost (per user) for this case.

Similar approach is exploited for the case with $M< K$. In this case, we
apply time sharing between delayed CSIT feedback with $\FBfracP'=0$ and
alternating CSIT feedback with $\FBfracP''= M/K$. In this case, the
first feedback strategy achieves $\dsum'=\Gamma$ by applying MAT scheme,
with the second strategy achieving $\dsum''=M$ by using alternating CSIT feedback manner.
As a result, for $\FBfracP \in [0, \frac{M}{K}]$ being the final feedback cost for this case, the final sum DoF is calculated as 
\begin{align*}
	\dsum &= \dsum'' +  (\dsum' - \dsum'')
        \frac{\FBfracP''-\FBfracP}{\FBfracP''-\FBfracP'} \\	
  &= (K - \frac{K \Gamma }{M})\FBfracP +\Gamma   
\end{align*}%
which completes the proof. 

\subsection{Proof of Proposition~\ref{lm:DoFM2KD} \label{sec:poof3}}

As shown in the Fig~\ref{fig:SumDoF_D_M2K}, the sum DoF performance has three regions: 
\begin{eqnarray}
  \dsum = \left\{
\begin{array}{l l}
  1+\frac{K}{2}\FBfracD, & \quad \FBfracD \in [0, \frac{2}{3K}] \\
	\frac{12}{11}+\frac{4K}{11} \FBfracD, & \quad \FBfracD \in [\frac{2}{3K}, \frac{9}{8K}] \\
	3/2, & \quad \FBfracD \in [\frac{9}{8K}, 1].
\end{array} \right. \nonumber
\end{eqnarray}

In the following, we will prove that the sum DoF $\dsum=1,\ \frac{4}{3},
\ \frac{3}{2} $ are achievable with $\FBfracD =0, \ \frac{2}{3K}, \ \frac{9}{8K}$, respectively. At the end, the whole DoF performance declared can be achievable by time sharing between those performance points.

The proposed scheme achieving $\dsum=\frac{4}{3}$ with $\FBfracD = \frac{2}{3K}$, is a modified version of the MAT scheme in \cite{MAT:11c}.
The new scheme has $K$ blocks, with each block consisting of three channel uses. In each block, 
four independent symbols are sent to two orderly selected users, which can be done with MAT scheme with each of two chosen user feeding back delayed CSIT in one channel use. 
As a result, $\dsum=\frac{4}{3}$ is achievable with $\FBfracD = \frac{2}{3K}$, using the fact that each of $K$ users needs to feed back delayed CSIT twice only in the whole communication (see Table~\ref{tab:scheme43d}).

\begin{table}
\caption{Summary of the achievability scheme for achieving
$\dsum=\frac{4}{3}$ with $\FBfracD = \frac{2}{3K}$.}
\begin{center}
\begin{tabular}{|c|c|c|c|c|c|c|}
  \hline
 block index          &  1             & 2               & 3        & $\cdots$                 & $K$  \\
   \hline
 No. of channel uses        &  3             & 3               & 3        & $\cdots$                 & 3  \\
   \hline
 Active users        &  user~1, 2     & user~2, 3        & user~3, 4 & $\cdots$                 & user~$K$, 1  \\
   \hline
	Delayed CSIT feedback   &  user~1: $1/3$  & user~2: $1/3$      & user~3: $1/3$           & $\cdots$   &   user~$K$: $1/3$  \\
	fraction in a block &  user~2: $1/3$  & user~3: $1/3$  & user~4: $1/3$ &             &   user~1: $1/3$\\
	                        &  the rest: 0  & the rest: 0 &the rest: 0 &            &  the rest: 0\\

   \hline
	Sum DoF           &  $4/3$          & $4/3$          & $4/3$         & $\cdots$   &   $4/3$  \\
	in a block    &    &   &  &            &   \\
	\hline
\end{tabular}
\end{center}
\label{tab:scheme43d}
\end{table}

Similarly, the proposed scheme achieving $\dsum=\frac{3}{2}$ with
$\FBfracD = \frac{9}{8K}$ has $K$ blocks, with each block consisting of 8 channel uses. In each block, 3 out of $K$ users are selected to communicate. In this case, $12$ independent symbols are sent to the chosen users during each block, which can be done with another MAT scheme with each of chosen users feeding back delayed CSIT in 3 channel uses. 
As a result, $\dsum=\frac{3}{2}$ is achievable with $\FBfracD = \frac{9}{8K}$, using the fact that each of $K$ users needs to feed back delayed CSIT  9 times only in the whole communication (see Table~\ref{tab:scheme32d}).

Finally, $\dsum=1$ is achievable without any CSIT. By now, we complete the proof.

\begin{table}
\caption{Summary of the achievability scheme for achieving
$\dsum=\frac{3}{2}$ with $\FBfracD = \frac{9}{8K}$.}
\begin{center}
\begin{tabular}{|c|c|c|c|c|c|c|}
  \hline
 block index          &  1             & 2               & 3        & $\cdots$                 & $K$  \\
   \hline
 No. of channel uses        &  8             & 8               & 8        & $\cdots$                 & 8  \\
   \hline
 Active users        &  user~1, 2, 3     & user~2, 3, 4        & user~3, 4, 5 & $\cdots$                 & user~$K$, 1, 2  \\
   \hline
	Delayed CSIT feedback   &  user~1: $3/8$  & user~2: $3/8$  & user~3: $3/8$ & $\cdots$    &   user~$K$: $3/8$  \\
	fraction in a block &  user~2: $3/8$  & user~3: $3/8$  & user~4: $3/8$ &             &   user~1: $3/8$\\
                          &  user~3: $3/8$  & user~4: $3/8$  & user~5: $3/8$ &             &   user~2: $3/8$\\
	                        &  the rest: 0    & the rest: 0    &the rest: 0    &             &  the rest: 0\\

   \hline
	Sum DoF           &  $3/2$          & $3/2$          & $3/2$         & $\cdots$   &   $3/2$  \\
	in a block    &    &   &  &            &   \\
	\hline
\end{tabular}
\end{center}
\label{tab:scheme32d}
\end{table}

\section{Conclusions} \label{sec:conclu}

This work considered the general multiuser MISO BC,
and established inner and outer bounds on the tradeoff between DoF performance and CSIT feedback quality, which are optimal for many cases. Those bounds, as well as some analysis, were provided with the aim of giving insights on how much CSIT feedback to achieve a certain DoF performance.

\section{Appendix - Proof details of Proposition~\ref{pro:entropydiff}} \label{sec:entropydiff}

In the following, we will prove Proposition~\ref{pro:entropydiff} used for the converse proof, as well as three lemmas to be used here.

\begin{lemma} \label{lemma:yi} \footnote{We note that Lemma~\ref{lemma:yi} is a slightly more general version of the result in \cite[Lemma~6]{YYGK:12}.} 
  Let $\Gm = \hat{\Gm} + \tilde{\Gm} \in \CC^{m\times m}$ where $\tilde{\Gm}$ has
  i.i.d.~$\mathcal{N}_c(0,1)$ entries, and $\tilde{\Gm}$ is independent
  of $\hat{\Gm}$. Then, we have 
  \begin{align}
    \E_{\tilde{G}} \bigl[ \log \det \left( \Gm^\H \Gm \right) \bigr] =
    \sum_{i=1}^{\tau} \log \bigl( \lambda_i(\hat{\Gm}^\H
    \hat{\Gm}) \bigr) + o(\log \snr) \label{eq:tmp777}
  \end{align}
  where $\lambda_i(\hat{\Gm}^\H \hat{\Gm})$ denotes the $i$ th largest
  eigenvalue of $\hat{\Gm}^\H\hat{\Gm}$; $\tau$ is the number of
  eigenvalues of $\hat{\Gm}^\H\hat{\Gm}$ that do not vanish with
  $\snr$, i.e., $\lambda_i(\hat{\Gm}^\H \hat{\Gm}) = o(1)$ when
  $\snr$ is large, $\forall\,i>\tau$. 
\end{lemma}

\begin{lemma} \label{lemma:bestpermutation}
  For $\Pm \in \CC^{m\times m}$ a permutation matrix and
  $\Am\in\CC^{m\times m}$, let $\Am \Pm =
  \Qm \Rm$ be the QR decomposition of the column permuted version of
  $\Am$. Then, there exist at least one permutation matrix $\Pm$ such that 
  \begin{align}
    r_{ii}^2 &\ge \frac{1}{{m-i+1}} \lambda_i(\Am^\H\Am), \quad i=1, \ldots,
    m \label{eq:tmp51}
  \end{align}%
  where as stated $\lambda_i(\Am^\H\Am)$ is the $i$ th largest eigenvalue of $\Am^\H\Am$;  $r_{ii}$ is the $i$~th diagonal elements of $\Rm$. 
\end{lemma}

\begin{lemma} \label{lemma:bestpermutation2}
  For any matrix $\Am\in\CC^{m\times m}$, there exists a column permuted
  version $\bar{\Am}$, such that
  \begin{align}
    \det( \bar{\Am}_{\mathcal{I}}^\H \bar{\Am}_{\mathcal{I}} ) &\ge
    m^{-|\mathcal{I}|} \prod_{i\in\mathcal{I}} \lambda_i(\Am^\H\Am), \quad
    \forall\,\mathcal{I} \subseteq \left\{ 1,\ldots,m \right\} 
  \end{align}%
  where $\bar{\Am}_{\mathcal{I}} = [A_{ji}:\ j\in\left\{ 1,\ldots,m
  \right\},\, i\in \mathcal{I}] \in \CC^{m\times
  |\mathcal{I}|}$ is the submatrix of $\Am$ formed by the columns
  with indices in $\mathcal{I}$.
\end{lemma}

\subsection{Proof of Lemma~\ref{lemma:yi} }
  
	Let us perform a singular value decomposition~(SVD) on the matrix
  $\hat{\Gm}$, i.e., $\hat{\Gm} = \Um \left[\begin{smallmatrix} \Dm_1 & \\ &
    \Dm_2 \end{smallmatrix}\right] \Vm^\H$ where $\Um, \Vm \in
    \mathbb{C}^{m\times m}$ are unitary matrices and $\Dm_1$ and
    $\Dm_2$ are $\tau' \times \tau'$ and $(m-\tau')\times (m-\tau')$ diagonal
    matrices of the singular values of $\hat{\Gm}$. 
    Without loss of generality, we assume that the $i$~th singluar value,
    $i=1,\ldots,m$, scales with $\snr$ as $\snr^{b_i}$, when $\snr$ is
    large. Moreover, the singular values in $\Dm_1$ are such that
    $b_i > 0$ and those in $\Dm_2$ verify $b_i\le 0$. 
    First, we have the following lower bound
  \begin{align}
    \MoveEqLeft[1]
    \E_{\tilde{G}} \bigl[ \log \det \left( \Gm^\H \Gm \right) 
    \bigr] \nonumber \\
    &=
\E_{\Mm} \biggl[ \log  \det \left( \left(  \left[\begin{smallmatrix} \Dm_1 & \\ &
      \Dm_2 \end{smallmatrix}\right] + \Mm \right)^\H 
      \left(  \left[\begin{smallmatrix} \Dm_1 & \\ &
      \Dm_2 \end{smallmatrix}\right] + \Mm \right)
      \right) \biggr] \\ 
      &\ge \E_{\Mm} \biggl[ \log  \det \left( \left(  \left[\begin{smallmatrix} \Dm_1 & \\ &
      0 \end{smallmatrix}\right] + \Mm \right)^\H 
      \left(  \left[\begin{smallmatrix} \Dm_1 & \\ &
      0 \end{smallmatrix}\right] + \Mm \right)
      \right) \biggr]  \label{eq:tmp239} \\
      &= \E_{\Mm}  \Bigl[  \log \bigl\lvert\det \left( \Dm_1 + \Mm_{11} \right) 
      \det \bigl( \Mm_{22} - \underbrace{\Mm_{21}\left( \Dm_1 + \Mm_{11} \right)^{-1}
      }_{\Bm} \Mm_{12} \bigr) \bigr \rvert^2 \Bigr] \label{eq:tmp232} \\ 
      &= \log \bigl\lvert\det \left( \Dm_1  \right) \bigr \rvert^2 +
      \E_{\Mm_{11}} \Bigl[ \log \bigl\lvert\det \left( I + \Dm_1^{-1} \Mm_{11}
      \right)  \bigr \rvert^2 \Bigr] + \E_{\Bm}
      \E_{\tilde{\Mm}} \bigl[ \log \det (\tilde{\Mm}^H (I + \Bm
      \Bm^H ) \tilde{\Mm}) \bigr] \label{eq:tmp236}\\
      &\ge  \log \bigl\lvert\det \left( \Dm_1 \right) \bigr \rvert^2 +
      \E_{\Mm_{11}} \Bigl[ \log \bigl\lvert\det \left( I + \Dm_1^{-1} \Mm_{11}
      \right)  \bigr \rvert^2 \Bigr] + \underbrace{\E_{\tilde{\Mm}}
      \bigl[ \log \det (\tilde{\Mm}^H  \tilde{\Mm}) \bigr] }_{(\ln2)^{-1}
      \sum_{l=0}^{m-\tau'-1}
      \psi(m-\tau'-l) = O(1)} \label{eq:tmp237}
  \end{align}
  where we define $\Mm \triangleq \Um^\H \tilde{\Gm} \Vm =
  \left[\begin{smallmatrix} \Mm_{11} & \Mm_{12} \\ \Mm_{21} & \Mm_{22}
  \end{smallmatrix}\right]$ with
  $\Mm_{11}\in\mathbb{C}^{\tau'\times\tau'}$, and remind that the entries of $\Mm$, thus
  of $\Mm_{ij}$, $i,j=1,2$, are also i.i.d.~$\mathcal{N}_c(0,1)$; \eqref{eq:tmp239}
  is from the fact that expectation of the log determinant of a
  non-central Wishart matrix is non-decreasing with in the
  ``line-of-sight'' component~\cite{kim2003log};  
  \eqref{eq:tmp232} is due to the identity $\det\left( \left[ \begin{smallmatrix} \Nm_{11} &
    \Nm_{12} \\ \Nm_{21} & \Nm_{22} \end{smallmatrix} \right] \right) =
    \det( \Nm_{11})\det( \Nm_{22} - \Nm_{21}
    \Nm_{11}^{-1} \Nm_{12})$ whenever $\Nm_{11}$ is
    square and invertible; in \eqref{eq:tmp236}, we notice that, given the matrix
    $\Bm\defeq\Mm_{21}\left( \Dm_1 + \Mm_{11} \right)^{-1}$, 
    the columns of $\Mm_{22} - \Bm \Mm_{12}$ are
    i.i.d.~$\mathcal{N}_c(0,I+\Bm \Bm^H)$, from which
    $\lvert\det(\Mm_{22}-\Bm \Mm_{12})\rvert^2$ is equivalent in
    distribution to $\det(\tilde{\Mm}^H (I+\Bm \Bm^H) \tilde{\Mm})$
    where $\tilde{\Mm} \in \mathbb{C}^{(m-\tau') \times (m-\tau')}$ has i.i.d.~$\mathcal{N}_c(0,1)$ entries; the
    last inequality is from $\tilde{\Mm}^H (I + \Bm \Bm^H ) \tilde{\Mm}
    \succeq \tilde{\Mm}^H\tilde{\Mm}$ and therefore $\det (\tilde{\Mm}^H
    (I + \Bm \Bm^H ) \tilde{\Mm}) \ge \det (\tilde{\Mm}^H \tilde{\Mm})$,
    $\forall\,\Bm$; the closed-form term in the last inequality is due
    to \cite{Muirhead} with $\psi(\cdot)$ being Euler's digamma
    function. In the following, we show that $\E \bigl[ \log
    \bigl\lvert\det \left( I + \Dm_1^{-1} \Mm_{11} \right)  \bigr
    \rvert^2 \bigr] \ge O(1)$ as well. To that end, we use the fact that the
    distribution of $\Mm_{11}$ is invariant to rotation, and so for
    $\Dm_1^{-1}\Mm_{11}$. Specifically, introducing
    $\theta\sim\textrm{Unif}(0,2\pi]$ that is independent of the rest of
    the random variables, we have
    \begin{align}
      \E_{\Mm_{11}} \Bigl[ \log \bigl\lvert\det \left( I + \Dm_1^{-1}
      \Mm_{11} \right)  \bigr \rvert^2 \Bigr] &= 
      \E_{\Mm_{11},\theta} \Bigl[ \log \bigl\lvert\det \left( I + \Dm_1^{-1} \Mm_{11} e^{j\theta} \right)  \bigr \rvert^2 \Bigr] \\
      &=\E_{\Mm_{11},\theta} \Bigl[ \log \bigl\lvert\det \left(
      e^{-j\theta} I + \Dm_1^{-1} \Mm_{11} \right)  \bigr \rvert^2 \Bigr] \\
      &= \sum_{i=1}^{\tau'} \E_{J} \E_{\theta} \bigl[ \log
      \lvert  e^{-j\theta} +
      \underbrace{\lambda_i(\Dm^{-1}\Mm_{11})}_{J_i} \rvert^2 \bigr] \label{eq:tmp344} \\
      &= \sum_{i=1}^{\tau'} \E_{J} \E_{\theta} [ \log  ( 1 +
      \lvert{J_{i}}\rvert^2 + 2  \lvert{J_{i}}\rvert
      \cos(\theta+\phi(J_i)) ) ] \label{eq:tmp349}\\
      &= \sum_{i=1}^{\tau'} \E_{J} \E_{\theta} [ \log  ( 1 +
      \lvert{J_{i}}\rvert^2 + 2  \lvert{J_{i}}\rvert \cos(\theta) ) ]
      \label{eq:tmp348}\\
      &\ge \sum_{i=1}^{\tau'} \bigl[ \E_{J} \bigl( \log  ( 1 +
      \lvert{J_{i}}\rvert^2 ) \bigr) - 1 \bigr]  \label{eq:tmp345} \\
      &\ge - \tau' \label{eq:tmp346}
    \end{align}%
    where the first equality is from the fact that $\Mm_{11}$ is
    equivalent to $\Mm_{11} e^{j\theta}$ as long as $\theta$ is
    independent of $\Mm_{11}$ and that $\Mm_{11}$ has independent
    circularly symmetric Gaussian entries; \eqref{eq:tmp344} is due to
    the characteristic polynomial of the matrix $-\Dm^{-1}
    \Mm_{11}$;  
    in \eqref{eq:tmp349} we define $\phi(J_i)$ the argument of
    $J_i$ that is independent of $\theta$; \eqref{eq:tmp348} is from the
    fact that
    $\mathrm{mod}(\theta+\phi)_{2\pi}\sim\textrm{Unif}(0,2\pi]$ and is
    independent of $\phi$, as long as $\theta\sim\textrm{Unif}(0,2\pi]$
    and is independent of $\phi$, also known as the Crypto
    Lemma~\cite{Forney:03}; 
    \eqref{eq:tmp345} is from the identity $\int_{0}^1 \log(a+b\cos(2 \pi t) )\, \mathrm{d} t =
    \log \frac{a + \sqrt{a^2-b^2}}{2} \ge \log(a) -1$, $\forall\,a\ge
    b>0$. Combining \eqref{eq:tmp237} and \eqref{eq:tmp346}, we have the
    lower bound 
    \begin{align}
      \E_{\tilde{G}} \bigl[ \log \det \left( \Gm^\H \Gm \right) 
    \bigr] \ge \log \bigl\lvert\det \left( \Dm_1 \right) \bigr
    \rvert^2 + O(1)  
    \end{align}%
    when $\snr$ is large. In fact, it has been shown
    that the $O(1)$ term here, sum of the $O(1)$ term in \eqref{eq:tmp237}
    and $-\tau'$ in \eqref{eq:tmp346}, does not depend on $\snr$ at all. 

    The next step is to derive an upper bound on $\E \bigl[\log
    \det \left( \Gm^\H \Gm \right)\bigr]$. Following Jensen's
    inequality, we have 
    \begin{align}
      \E_{\tilde{G}} \bigl[ \log \det ( \Gm^\H \Gm )\bigr] &\le  
      \log \det \bigl( \E_{\tilde{G}} [ \Gm^\H \Gm ] \bigr) \\
      &= \log \det \left(\left[\begin{smallmatrix} \Dm_1^2 & \\ &
        \Dm_2^2 \end{smallmatrix}\right] + \E [\Mm^\H \Mm] \right) \\ 
      &= \log \lvert \det \left( \Dm_1 \right) \rvert^2 + \underbrace{\log \det
      \left( I + m \Dm_1^{-2} \right)}_{o(1)} + \underbrace{\log \det
      \left( m I + \Dm_2^2 \right)}_{o(\log\snr)} \\
      &= \log \lvert \det \left( \Dm_1 \right) \rvert^2 + o(\log\snr)
    \end{align}%
    Putting the lower and upper bounds together, we have  
      $\mathbb{E} \bigl[ \log \det \left( \Gm^\H \Gm \right) \bigr] 
      = \log \lvert \det \left( \Dm_1 \right) \rvert^2 + o(\log\snr)$. 
    Finally, note that, since $\lambda_i(\hat{\Gm}^\H
    \hat{\Gm}) \doteq \snr^0$, $i=\tau'+1,\ldots,\tau$, we have 
      \begin{align}
        \log \bigl\lvert\det \left( \Dm_1  \right)  \bigr \rvert^2  &= 
\sum_{i=1}^{\tau'} \log \bigl( \lambda_i(\hat{\Gm}^\H \hat{\Gm}) \bigr) \\
&=\sum_{i=1}^{\tau} \log \bigl( \lambda_i(\hat{\Gm}^\H
\hat{\Gm}) \bigr) - \sum_{i=\tau'+1}^{\tau} \log \bigl(
\lambda_i(\hat{\Gm}^\H \hat{\Gm}) \bigr) \\
&=\sum_{i=1}^{\tau} \log \bigl( \lambda_i(\hat{\Gm}^\H
\hat{\Gm}) \bigr) + o(\log\snr)      
\end{align}%
from which the proof is complete.

\subsection{Proof of Lemma~\ref{lemma:bestpermutation}}
  The existence is proved by construction. Let $\av_j$, $j=1,\ldots,m$, be the $j$~th
  column of $\Am$. We define $j_1^*$ as the index of the column that has
  the largest Euclidean norm, i.e., 
  \begin{align}
    j_1^* &= \arg\max_{j=1,\ldots,m} \|\av_j\|. 
  \end{align}%
  Swapping the $j_1^*$ and the first column, and denoting $\Am_1 = \Am$, we have
  \begin{align}
    \Bm_1 &\triangleq \Am_1 \Tm_{1,j_1^*}
  \end{align}%
  where $\Tm_{ij}\in \CC^{m\times m}$ denotes the permutation matrix that swaps the
  $i$~th and $j$~th columns. Now, let $\Um_1\in \CC^{m\times m}$ be any
  unitary matrix such that the first column is aligned with the first
  column of $\Bm_1$, i.e., equal to
  $\frac{\av_{j_1^*}}{\|\av_{j_1^*}\|}$. 
  Then, we can construct a block-upper-triangular matrix $\Rm_1 = \Um_1^\H
  \Bm_1 = \Um_1^\H \Am_1 \Tm_{1,j_1^*}$ with the following form 
  \begin{align}
    \Rm_1 &= \begin{bmatrix} r_{11} & * \\ \mathbf{0}_{(m-1)\times 1} & \Am_2 \end{bmatrix}
  \end{align}%
  where it is readily shown that
  \begin{align}
    r_{11}^2 &= \|\av_{j_1^*}\|^2 \\
    &\ge \frac{1}{m} || \Am_1 ||^2_F \\
    &\ge \frac{1}{m} \lambda_1(\Am_1^\H\Am_1).
  \end{align}%
  Repeating the same procedure on $\Am_2$, we will have 
  $\Rm_2 = \Um_2^\H \Bm_2 = \Um_2^\H \Am_2 \Tm_{2,j_2^*}$ where all the
  involved matrices are similarly defined as above except for the
  reduced dimension $(m-1)\times(m-1)$ and 
   \begin{align}
    \Rm_2 &= \begin{bmatrix} r_{22} & * \\ \mathbf{0}_{(m-2)\times 1} & \Am_3 \end{bmatrix}
  \end{align}%
  where it is readily shown that
  \begin{align}
    r_{22}^2 &\ge \frac{1}{m-1} \lambda_1(\Am_2^\H\Am_2) \\
    &\ge \frac{1}{m-1} \lambda_2(\Am_1^\H\Am_1).
  \end{align}%
  Here, the last inequality is from the fact that, for any matrix
  $\Cm$ and a submatrix $\Cm_k$ by removing $k$ rows or columns, we
  have~\cite[Corollary 3.1.3]{HJ:91}
  \begin{align}
    \lambda_i(\Cm_k^\H\Cm_k)\ge\lambda_{i+k}(\Cm^\H\Cm)
  \end{align}%
  where we recall that $\lambda_i$ is the $i$~th largest eigenvalue. 
  Let us continue the procedure on $\Am_3$ and so on. At the end, we
  will have all the $\left\{ \Um_i \right\}$ and $\left\{ \Tm_{i,j_i^*}
  \right\}$ such that 
  \begin{align}
    \underbrace{\left[\begin{smallmatrix} I_{m-1} & \\ & \Um_m^\H  \end{smallmatrix}
      \right]\cdots\left[\begin{smallmatrix} I_2 & \\ & \Um_3^\H  \end{smallmatrix}
      \right]\left[\begin{smallmatrix} 1 & \\ & \Um_2^\H  \end{smallmatrix}
        \right] \Um_1^\H}_{\Qm^\H} \Am \underbrace{\Tm_{1,j_1^*} \left[\begin{smallmatrix} 1 &
        \\ & \Tm_{2,j_2^*}  \end{smallmatrix} \right]
        \left[\begin{smallmatrix} I_2 &
        \\ & \Tm_{3,j_3^*}  \end{smallmatrix} \right] \cdots
        \left[\begin{smallmatrix} I_{m-1} &
          \\ & \Tm_{m,j_m^*}  \end{smallmatrix} \right]}_{\Pm}
        &= \underbrace{\begin{bmatrix} r_{11} & * & * & * \\  & r_{22} & * & *  \\ & & \ddots
          & \vdots \\ & & & r_{mm} \end{bmatrix}}_{\Rm}
  \end{align}%
  where it is obvious that $\Pm$ is a permutation matrix and $\Qm$ is
  unitary. The proof is thus completed.

\subsection{Proof of Lemma~\ref{lemma:bestpermutation2}}
Let $\bar{\Am} \triangleq \Am \Pm = \Qm \Rm$ with $\Pm$ a permutation matrix such that
\eqref{eq:tmp51} holds. Then, we have 
\begin{align}
  \det( \bar{\Am}_{\mathcal{I}}^\H \bar{\Am}_{\mathcal{I}} ) &= 
  \det( {\Rm}_{\mathcal{I}}^\H \Qm^\H \Qm {\Rm}_{\mathcal{I}} ) \\
  &= \det( {\Rm}_{\mathcal{I}}^\H  {\Rm}_{\mathcal{I}} ) \\
  &\ge \det( {\Rm}_{\mathcal{I}\mathcal{I}}^\H  {\Rm}_{\mathcal{I}\mathcal{I}} ) \\
  &= \prod_{i\in\mathcal{I}} r_{ii}^2 \\ 
  &\ge m^{-|\mathcal{I}|} \prod_{i\in\mathcal{I}} \lambda_i(\Am^\H\Am)
\end{align}%
where the first inequality results from the Cauchy-Binet formula, and the last inequality is due to Lemma~\ref{lemma:bestpermutation}.

\subsection{Proof of Proposition~\ref{pro:entropydiff}}

The inequality \eqref{eq:tmp2} is trivial when $m \ge l \ge M$, i.e.,
$l'=m'=M$. From the chain rule $h(\yv_m \cond U,\hat{H},\tilde{H}) =
h(\yv_l \cond U,\hat{H},\tilde{H} ) +
h(y_{l+1},\ldots,y_{m} \cond \yv_l,\hat{H},\tilde{H} ) = h(\yv_l\cond
U,\hat{H},\tilde{H} ) + o(\log \snr)$, since with $l\ge M$, the
observations $y_{l+1},\ldots,y_{m}$ can be
represented as a linear combination of $\yv_l$, up to the noise error.
In the following, we focus on the case $l \le M$.

First of all, let us write
\begin{multline}
   h(\yv_{m}|U,\hat{H},\tilde{H}) - \mu\, h(\yv_l|U,\hat{H},\tilde{H})
   \\ = \E_{\hat{H}}\Bigl [ \E_{\tilde{H}}[h(\Hm_m \xv + \zv_m \cond
   U,\hat{H}=\hat{\Hm},\tilde{H} = \tilde{\Hm})]
    - \mu \,
   \E_{\tilde{H}}[h(\Hm_l \xv + \zv_l \cond U,\hat{H}=\hat{\Hm},\tilde{H} =
   \tilde{\Hm})] \Bigr ] \label{eq:tmp129}
\end{multline}%
In the following, we focus on the term inside the expection over
$\hat{H}$ in \eqref{eq:tmp129},
i.e., for a given realization of $\hat{\Hm}$. 
Since $\yv_l$ is a degraded version of $\yv_m$, we can apply the results
in \cite[Corollary~4]{WLS+:09} and obtain the optimality
of Gaussian input, i.e., 
\begin{align}
  &\max_{\stackrel{p_{X|U\hat{H}}:}{\E[\text{tr}(X X^\H)]\le  \snr}}
  \E_{\tilde{H}} \bigl[ h(\yv_{m}|U,\hat{H}=\hat{\Hm},\tilde{H}=
  \tilde{\Hm})\bigr] -
  \mu\,\E_{\tilde{H}} \bigl[
   h(\yv_l|U,\hat{H}=\hat{\Hm},\tilde{H}= \tilde{\Hm}) \bigr] \nonumber \\
   &= \max_{\Psim\succeq 0: \text{tr}(\Psim)\le  \snr} \E_{\tilde{H}}
   \bigl[
   \log \det \left( I + \Hm_m \Psim \Hm_m^\H \right) \bigr]
   - \mu\,\E_{\tilde{H}} \bigl[ \log \det \left( I + \Hm_l \Psim \Hm_l^\H \right)
   \bigr] \label{eq:tmp3}
\end{align}%
for any $\mu \ge 1$. The next step is to upper bound the right hand
side~(RHS) of \eqref{eq:tmp3}.  

Next, let $\Psim = \Vm \Lambdam \Vm^\H$ be the eigenvalue decomposition of the
covariance matrix $\Psim$ where $\Lambdam$ is a diagonal matrix and $\Vm$ is unitary. 
Note that it is without loss of generality to assume that all
eigenvalues of $\Psim$ are strictly positive, i.e., $\lambda_i (\Psim) \ge c >0$,
$\forall i$, in the sense that 
\begin{align}
  \log \det \left( I + \Hm\Psim \Hm^\H \right) 
  \le \log \det \left( I + \Hm (cI + \Psim) \Hm^\H \right) 
   \le \log \det \left( I + \Hm\Psim \Hm^\H \right) + \log \det \left( I
   + c \Hm \Hm^\H \right). 
\end{align}
In other words, a constant lift of the eigenvalues of $\Psim$ does not
have any impact on the high snr behavior. 
This regularization will however simplify the analysis.  
The following is an upper bound for the first term in the RHS of
\eqref{eq:tmp3}.
\begin{align}
  \E_{\tilde{H}} \Bigl[ \log \det \bigl( I + \Hm_m \Psim \Hm_m^\H \bigr)
  \Bigr] 
  &= \E_{\tilde{H}} \bigl[ \log \det \left( I_M +   \Psim^{\frac{1}{2}}
  \Hm_m^\H \Hm_m \Psim^{\frac{1}{2}} \right) \bigr] \label{eq:tmp47}\\
  &\le \E_{\tilde{H}} \biggl[ \log \det \left( I_M + \Psim^{\frac{1}{2}}
  \Um^H \left[ \begin{smallmatrix} \|\Hm_m\|^2_F
    I_{m'} & \\ & 0 \end{smallmatrix} \right] \Um \Psim^{\frac{1}{2}} \right) \biggr] \label{eq:tmp45}\\
    &= \E_{\tilde{H}} \Bigl[ \log \det \bigl( I_{m'} +
    \|\Hm_m\|^2_F\, \tilde{\Psim} \bigr) \Bigr] \label{eq:tmp46}\\
    &= \E_{\tilde{H}} \bigl[ \log \det \bigl( \tilde{\Psim} \bigr)
    \bigr] + \E_{\tilde{H}}
    \Bigl[ \log \det \bigl( \tilde{\Psim}^{-1} + \|\Hm_m\|^2_F I \bigr)
    \Bigr] \label{eq:tmp48}\\
    &\le \sum_{i=1}^{m'} \!\! \log \lambda_i(\Psim) + \underbrace{\log \det
  \bigl( (c^{-1} + m + \|\hat{\Hm}_m\|^2_F) I \bigr)}_{o(\log\snr)} \label{eq:tmp43} \\
  &\le \log \det(\Lambdam) + o(\log \snr) \label{eq:tmp42}
\end{align}%
where $\Psim^{\frac{1}{2}}$ is such that
$\bigl(\Psim^{\frac{1}{2}}\bigr)^2 = \Psim$; \eqref{eq:tmp45} is due
to fact that $\Hm_m^H \Hm_m \preceq \Um^H \left[ \begin{smallmatrix}
  \|\Hm_m \|^2_F I_{m'} & \\ & 0 \end{smallmatrix} \right] \Um$ with
    $\Um$ being the matrix of eigenvectors of $\Hm_m^H \Hm_m$ and $\|\Hm_m\|_F$ being the Frobenius norm of $\Hm_m$, where $m'\defeq \min\{m,M\}$;  
		in \eqref{eq:tmp46}, we define $\tilde{\Psim}$ as the $m'\times m'$ upper left block of $\Um \Psim \Um^\H$; the first term in 
    \eqref{eq:tmp43} is due to $\det(
    \tilde{\Psim} ) = \prod_{i=1}^{m'}
    \lambda_i(\tilde{\Psim}) \le \prod_{i=1}^{m'}
    \lambda_i(\Um \Psim \Um^H) = \prod_{i=1}^{m'}
    \lambda_i(\Psim)$; 
the second term in \eqref{eq:tmp43}  is from Jensen's inequality and using the fact
that $\Psim^{-1} \preceq c^{-1} I_M$ by assumption and that
$\mathbb{E}_{\tilde{H}} (\Hm_m^H \Hm_m) = \sum_{k=1}^m \sigma_k^2 I_M + \hat{\Hm}_m^H
\hat{\Hm}_m
\preceq m I + \hat{\Hm}_m^H \hat{\Hm}_m$; the last inequality is from
the assumption that every eigenvalue of $\Psim$ is lower-bounded by
some constant~$c>0$ independent of $\snr$. Now, we need to lower bound the
second expectation in the RHS of \eqref{eq:tmp3}. To this end, let us write 
\begin{align}
\det \left( I_l + \Hm_l \Psim \Hm_l^\H \right) 
   &= \det \left( I_l + \Hm_l \Vm \Lambdam \Vm^\H \Hm_l^\H \right)
   \label{eq:tmp10}\\ 
   &= \det \left( I_{M} + \Lambdam \Vm^\H \Hm_l^\H \Hm_l \Vm \right)
   \label{eq:tmp11}\\
   &= \det \left( I_{M} + \Lambdam \Phim^\H \Sigmam^2 \Phim \right) \label{eq:tmp12}\\
   &= 1+\sum_{\mathcal{I} \subseteq \left\{ 1,\ldots,M \right\}, 
   \mathcal{I}\ne \emptyset}
   \!\! \det(\Lambdam_{\mathcal{I}\mathcal{I}})
   \det(\Phim_{\mathcal{I}}^\H \Sigmam^2  \Phim_{\mathcal{I}}) \label{eq:tmp13}\\
   &\ge \det(\Sigmam^2) \sum_{j = 1}^{M} \det(\Lambdam_{\mathcal{I}_j\mathcal{I}_j}) \det(\Phim_{\mathcal{I}_j}^\H \Phim_{\mathcal{I}_j}) \label{eq:tmp14}\\ 
   &\ge M  \det(\Sigmam^2) \left( \prod_{j = 1}^{M} \left( \det(\Lambdam_{\mathcal{I}_j\mathcal{I}_j}) \det(\Phim_{\mathcal{I}_j}^\H \Phim_{\mathcal{I}_j}) \right) \right)^{\frac{1}{M}} \label{eq:tmp15}\\ 
   &= M  \det(\Sigmam^2) \det(\Lambdam)^{\frac{l}{M}} \left( \prod_{j =
   1}^{M} \det(\Phim_{\mathcal{I}_j}^\H \Phim_{\mathcal{I}_j})
   \right)^{\frac{1}{M}}  \label{eq:tmp8}
\end{align}%
where \eqref{eq:tmp11} is an application of the identity
$\det(I + \Am \Bm) = \det(I + \Bm \Am)$; 
in \eqref{eq:tmp12}, we define 
\[
\Sigmam \triangleq \diag(\sigma_1, \ldots, \sigma_l), \ \Phim \triangleq
\Sigmam^{-1}\Hm_l \Vm, \ \text{and} \  
\hat{\Phim} \triangleq \Sigmam^{-1}\hat{\Hm_l} \Vm;
\] 
in \eqref{eq:tmp13}, we
define 
$\Phim_{\mathcal{I}} \triangleq [\Phi_{ji}:\ j=1,\ldots,l,\
i\in\mathcal{I} ]\in\CC^{l\times |\mathcal{I}|}$ as the submatrix of
$\Phim$ with columns indexed in $\mathcal{I}$ and $\Lambdam_{\mathcal{I}\mathcal{I}} = [\Lambda_{ji}:\ i,j\in\mathcal{I}]\in\CC^{|\mathcal{I}|\times
|\mathcal{I}|}$, with $\mathcal{I}$ denoting a nonempty set; the equality \eqref{eq:tmp13} is an application of the
identity \cite{Aitken:54d} \[\det(I + \Am) = 1+\!\sum_{\mathcal{I}\subseteq\left\{ 1,\ldots,M
\right\}, \mathcal{I}\ne\emptyset} \!\! \det(\Am_{\mathcal{I}\mathcal{I}})\] for any $\Am\in
\CC^{M\times M}$; 
in \eqref{eq:tmp14}, we define 
$\mathcal{I}_1, \ldots, \mathcal{I}_M$ as the so-called sliding
window of indices
\begin{align}
  \mathcal{I}_1 &\triangleq \{ 1,2,\cdots,l\}, \ \mathcal{I}_2
  \triangleq \{ 2,3,\cdots,l,l+1\}, \ \cdots, \  \mathcal{I}_M
  \triangleq \{ M,1,2,\cdots,l-1\} \\ 
  \text{i.e.}, \  \mathcal{I}_j &\triangleq \left\{ \mathrm{mod}(j+i-1)_M + 1: \ i=0,1,\cdots,l-1 \right\}, \ j=1,2,\cdots, M
  \end{align}%
  with $\mathrm{mod}(x)_M$ being the modulo operator; 
\eqref{eq:tmp15} is from the fact that arithmetic mean
is not smaller than geometric mean; in \eqref{eq:tmp8}, we use the fact
that $\prod_{j = 1}^{M}  \det(\Lambdam_{\mathcal{I}_j\mathcal{I}_j}) =
\det(\Lambdam)^{l}$.
 
Without loss of generality, we assume that the $M$ columns of $\Hm_l \Vm$ are
ordered in such a way that 1) the first $l$ columns are linearly
independent, i.e., $\hat{\Phim}_{\mathcal{I}_1}$ has full rank, 
and 2) $\Am = \hat{\Phim}_{\mathcal{I}_1}$ satisfies
Lemma~\ref{lemma:bestpermutation2}. Note that the former condition can
almost always be
satisfied since $\rank(\hat{\Phim}) = l$ almost surely. Hence, we have
\begin{align}
  \E_{\tilde{H}} \bigl[ \log \det(\Phim_{\mathcal{I}_j}^\H
  \Phim_{\mathcal{I}_j}) \bigr]
  &= 
  \sum_{i=1}^{\rank(\hat{\Phim}_{\mathcal{I}_j})} \!\! \log
  \bigl(\lambda_i(\hat{\Phim}_{\mathcal{I}_j}^\H\hat{\Phim}_{\mathcal{I}_j})\bigr)  + o(\log \snr)
  \label{eq:tmp21}\\
  &\ge
  \sum_{i=1}^{\rank(\hat{\Phim}_{\mathcal{I}_j\bigcap\mathcal{I}_1})}
  \!\!\!
  \log\bigl(\lambda_i(\hat{\Phim}_{\mathcal{I}_j}^\H\hat{\Phim}_{\mathcal{I}_j})\bigr)  + o(\log \snr) \label{eq:tmp22}\\
  &\ge
  \sum_{i=1}^{\rank(\hat{\Phim}_{\mathcal{I}_j\bigcap\mathcal{I}_1})}
  \!\!\! \log
  \bigl(\lambda_i(\hat{\Phim}_{\mathcal{I}_j\bigcap\mathcal{I}_1}^\H\hat{\Phim}_{\mathcal{I}_j\bigcap\mathcal{I}_1})\bigr)  + o(\log \snr) \label{eq:tmp23}\\
  &= \log \det(\hat{\Phim}^\H_{\mathcal{I}_j\bigcap\mathcal{I}_1}\hat{\Phim}_{\mathcal{I}_j\bigcap\mathcal{I}_1})  + o(\log \snr) \\
  &\ge \log \!\! \prod_{i\in\mathcal{I}_j\bigcap\mathcal{I}_1}\!\!\!\lambda_i(\hat{\Phim}^\H\hat{\Phim})  + o(\log \snr) 
\end{align}%
where \eqref{eq:tmp21} is from Lemma~\ref{lemma:yi} by noticing that
$\Phim_{\mathcal{I}_j} =
\hat{\Phim}_{\mathcal{I}_j}+\tilde{\Phim}_{\mathcal{I}_j}$ with
the entries of $\tilde{\Phim}_{\mathcal{I}_j} \triangleq \Sigmam^{-1}\tilde{\Hm_l} \Vm$
being i.i.d.~$\mathcal{N}_c(0,1)$; \eqref{eq:tmp22} is
from the fact that $\rank(\hat{\Phim}_{\mathcal{I}_j}) \ge
\rank(\hat{\Phim}_{\mathcal{I}_j\bigcap\mathcal{I}_1})$;
\eqref{eq:tmp23} is due to
$\lambda_i(\hat{\Phim}_{\mathcal{I}_j}^\H\hat{\Phim}_{\mathcal{I}_j}) \ge
\lambda_i(\hat{\Phim}_{\mathcal{I}_j\bigcap\mathcal{I}_1}^\H\hat{\Phim}_{\mathcal{I}_j\bigcap\mathcal{I}_1})$ where we
recall that $\lambda_i(\Am^\H\Am)$ is defined as the $i$~th largest eigenvalue
of $\Am^\H \Am$; and the last inequality is due to
Lemma~\ref{lemma:bestpermutation2}. Summing over all $j$, we have
\begin{align}
  \sum_{j=1}^M \E_{\tilde{H}} \bigl[ \log \det(\Phim_{\mathcal{I}_j}^\H
  \Phim_{\mathcal{I}_j}) \bigr]  
  &\ge \log \Biggl( \prod_{j=1}^M
  \prod_{i\in\mathcal{I}_j\bigcap\mathcal{I}_1}\!\!\!\lambda_i(\hat{\Phim}^\H\hat{\Phim})\Biggr) + o(\log \snr) \\
  &= \log \Biggl( \Biggl(
  \prod_{i\in\mathcal{I}_1}\lambda_i(\hat{\Phim}^\H\hat{\Phim})\Biggr)^{\!\!l}
  \Biggr) + o(\log \snr) \\
  &\ge l \log 
  \prod_{i\in\mathcal{I}_1}\lambda_i(\hat{\Phim}_{\mathcal{I}_1}^\H\hat{\Phim}_{\mathcal{I}_1})
   + o(\log \snr) \label{eq:tmp31} \\
  &= l \log \det \left( \hat{\Phim}^\H_{\mathcal{I}_1}\hat{\Phim}_{\mathcal{I}_1}\right)  + o(\log \snr) \\
  &= -l \log \det \left( \Sigmam^2 \right)  + o(\log \snr)
  \label{eq:tmp4}
\end{align}
where \eqref{eq:tmp31} is due to $\lambda_i(\hat{\Phim}^\H\hat{\Phim}) \ge
\lambda_i(\hat{\Phim}_{\mathcal{I}_1}^\H\hat{\Phim}_{\mathcal{I}_1})$, $\forall\, i=1,\ldots,l$;
the last equality is from the fact that $\hat{\Phim}_{\mathcal{I}_1} =
\Sigmam^{-1} \hat{\Hm}_l \Vm_{\mathcal{I}_1}$ and that $\hat{\Hm}_l
\Vm_{\mathcal{I}_1}$ has full rank by construction. From \eqref{eq:tmp8} and \eqref{eq:tmp4}, we obtain
\begin{equation}
  \E_{\tilde{H}}\bigl[ \log  \det \left( I_l + \Hm_l \Psim \Hm_l^\H \right)
  \bigr]  \ge \frac{l}{M} \log \det(\Lambdam) + \frac{M-l}{M} \log
  \det(\Sigmam^2) + o(\log \snr)
\end{equation}%
and finally
\begin{equation}
  \E_{\tilde{H}} \bigl[ \log \det \left( I_m + \Hm_m \Psim \Hm_m^\H
  \right) \bigr]
   - \frac{M}{l} \E_{\tilde{H}} \bigl[ \log \det \left( I_l +
  \Hm_l \Psim \Hm_l^\H \right)\bigr] \le   - \frac{M-l}{l} \log
  \det(\Sigmam^2) + o(\log \snr).  \label{eq:tmp41}
\end{equation}%
When $m<M$, the above bound \eqref{eq:tmp41} is not tight. However, we
can show that, in this case, \eqref{eq:tmp41} still holds when we
replace $M$ with $m$. To see this, let us define $\Lambdam' \triangleq
\diag(\lambda_1,\ldots,\lambda_m)$. First, note that when $m< M$,
\eqref{eq:tmp42} holds if we replace $\Lambdam$ with $\Lambdam'$ on
the RHS. Then, the RHS of \eqref{eq:tmp10} becomes a lower bound if we
replace $\Lambdam$ with $\Lambdam'$ and $\Vm$ with
$\Vm'\in\CC^{M\times m}$, the first $m$ columns of $\Vm$. From then on,
every step holds with $M$ replaced by $m$. \eqref{eq:tmp41} thus follows
with $M$ replaced by $m$. By taking the expectation on both sides of
\eqref{eq:tmp41} over $\hat{\Hm}$ and plugging it into \eqref{eq:tmp129}, we complete the proof of \eqref{eq:tmp2}.



\end{document}